\documentclass[onecolumn,draftclsnofoot]{IEEEtran}
\usepackage{amsmath}
\usepackage{graphicx}
\usepackage{amsthm}
\usepackage{float}
\usepackage{verbatim}
\usepackage{float}
\usepackage{epstopdf}
\usepackage{amssymb}
\usepackage{amsfonts}
\usepackage{color}
\usepackage{subcaption}
\usepackage{caption}

\begin{document}
\title{Constant Envelope Signaling in MIMO Channels}
\author{{Borzoo Rassouli and Bruno Clerckx}
\thanks{Borzoo Rassouli is with the Communication and Signal Processing group of Department of Electrical and Electronics,
Imperial College London, United Kingdom. email: b.rassouli12@imperial.ac.uk}
\thanks{Bruno Clerckx is with the Communication and Signal Processing group of Department of Electrical and Electronics,
Imperial College London and the School of Electrical Engineering, Korea University, Korea. email: b.clerckx@imperial.ac.uk}
\thanks{This work was partially supported by the Seventh Framework Programme for Research of the European Commission under grant number HARP-318489.}}
\maketitle
\begin{abstract}
%\boldmath
The capacity of the point-to-point vector Gaussian channel under the peak power constraint is not known in general. This paper considers a simpler scenario in which the input signal vector is forced to have a constant envelope (or norm). The capacity-achieving distribution for the non-identity $2\times 2$ MIMO channel when the input vector lies on a circle in $\mathbb{R}^2$ is obtained and is shown to have a finite number of mass points on the circle. Subsequently, it is shown that the degrees of freedom (DoF) of a full-rank $n$ by $n$ channel with constant envelope signaling is $n-1$ and it can be achieved by a uniform distribution over the surface of the hypersphere whose radius is defined by the constant envelope. Finally, for the 2 by 2 channel, the power allocation scheme of the constant envelope signaling is compared with that of the conventional case, in which the constraint is on the average transmitted power. It is observed that when the condition number of the channel is close to one, both schemes have a similar trend while this is not the case as the condition number grows.
\end{abstract}

\section{Introduction}\label{Intro}
In \cite{Smith}, the capacity of the point-to-point scalar Gaussian channel under the peak and average power constraints was investigated and it was shown that the capacity-achieving distribution has a probability mass function with a finite number of mass points. Shamai and Bar-David gave a full account on the capacity of a quadrature Gaussian channel under the aforementioned constraints in \cite{Shamai} and proved that the optimal input distribution has a discrete amplitude and a uniform independent phase. Even without a peak power constraint, this discreteness in the optimal input distribution was shown in \cite{Abou} to be true for the Rayleigh-fading channel when no channel state information (CSI) is assumed either at the receiver or the transmitter. Following this work, the authors in \cite{Katz} and \cite{Gursoy} investigated the capacity of noncoherent AWGN and Rician-fading channels, respectively. In \cite{Tchamkerten}, a point to point real scalar channel is considered in which sufficient conditions for the additive noise are provided such that the support of the optimal bounded input has a finite number of mass points. These sufficient conditions are also useful in multi-user settings as shown in \cite{khandani} for the MAC channel under bounded inputs.

The capacity of the Gaussian MIMO with identity channel under the peak and average power constraints is shown in \cite{Borzoo} where the support of the optimal input
distribution is a finite set of hyper-spheres with mutual independent phases and amplitude in the spherical domain. However, the capacity of the general point-to-point Gaussian MIMO channel under the peak power constraint is an open problem. In this paper, we address a simpler problem in which the input is forced to have a constant envelope (i.e., for any codeword $\mathbf x^n(m)$ where $m$ denotes the message index, instead of the peak power constraint which is equivalent to $\|\mathbf x_i(m)\|\leq R\ ,\ \forall i\in[1:n]$, a stronger condition, which is $\|\mathbf x_i(m)\|= R\ ,\ \forall i\in[1:n]$, must be satisfied). A 2 by 2 non-identity channel matrix is considered. The capacity of this channel under constant-norm inputs is obtained and it is shown that the capacity achieving distribution has a finite number of mass points on the circle defined by the constraint. Although the capacity does not have a closed form solution, lower and upper bounds can be obtained for it which are sufficient to give the optimal degrees of freedom (DoF). As a result, it is shown that the degrees of freedom (DoF) of a full-rank $n$ by $n$ channel with constant envelope signaling is $n-1$ and it can be achieved by a uniform distribution over the surface of the hypersphere whose radius is defined by the constant envelope. Finally, for the 2 by 2 channel, the constant envelope signaling is compared to the conventional case which has only the average power constraint. It is shown that when the condition number of the channel is close to one (i.e., the channel is ill-conditioned), the optimal power allocation scheme of the constant envelope signaling has a similar behavior to that of the conventional case, which is the water-filling algorithm. However, as the condition number grows, the schemes show completely different trends. The criterion for this comparison is the power level at which the weaker channel starts to be allocated non-zero power which is evaluated in the numerical results.

The steps for proving the finiteness of the support of the optimal input is similar to that in \cite{Smith} which is based on contradiction. More precisely, first, it is assumed that the optimal input has an infinite number of mass points. By using some tools in real and complex analysis, this assumption leads to an equality (involving a probability density function) which must be satisfied on a set. The last part of the proof is showing that this equality does not hold, and therefore disproving the first assumption of an infinite number of points for the optimal input distribution. In \cite{Smith} and \cite{Shamai} this contradiction is obtained by directly solving for the probability density function (by means of Fourier and Laplace transforms) and showing that either it is not a legitimate pdf or it cannot be induced by the input. Hermite polynomials and its properties were used in \cite{Fahs} to solve for the probability density function and get the contradiction. The application of these methods and solving for the pdf is not straightforward for the problem considered in this paper. Therefore, knowing that the right hand side of the aforementioned equality is a constant, we obtain the contradiction by showing that the left hand side of this equality can become unbounded with its parameter.

The paper is organized as follows. Section \ref{sm} explains the system model under consideration. Section \ref{th} states the main result of this paper through a theorem whose detailed proof is given in section \ref{pth}. The asymptotic behavior of the capacity-achieving input distribution for small values of SNR along with the degrees of freedom under constant envelope signaling are presented in section \ref{asymbeh} . In section \ref{polar}, the problem is analyzed in the polar coordinates and the notations of this section will be used in section \ref{nr} which shows the numerical results. The paper is concluded in section \ref{conc}.

%The paper is organized as follows.

%----------------------------------------------------------------------------------------
\section{system model}\label{sm}
We consider a $2\times2$ discrete-time memoryless vector Gaussian channel given by
\begin{equation}\label{4e1}
    \mathbf{Y}_i = \mathbf{H}\mathbf{X}_i + \mathbf{W}_i
\end{equation}
in which $i$ denotes the channel use. $\mathbf{H}=\mbox{diag}\{\lambda,1\}$ ($|\lambda|\neq 1$) is the deterministic channel matrix and $\{\mathbf{W}_i\}$ is an i.i.d. noise vector process with $\mathbf{W}_i\sim N(\mathbf 0,\mathbf{I}_2)$ (and independent of $\mathbf{X}_i$) for every transmission $i\in[1:n]$. The assumption of $|\lambda|\neq 1$ is to exclude the identity channel matrix for which the capacity-achieving distribution under a fixed transmission power is already known in \cite{Wyner} (i.e., the optimal input has uniform phase on the circle defined by the constant norm). It can be easily verified that it is sufficient to consider only the case $\lambda >1$. \footnote{This can be justified by a simple normalization and symmetry of the noise.}

The capacity of this channel under a fixed transmission power (i.e., constant norm) is
\begin{equation}\label{4e2}
    C(R) = \sup_{F_{\mathbf{X}}(\mathbf{x}):\|\mathbf{X}\|\stackrel{a.s.}=R} I(\mathbf{X};\mathbf{Y})=\sup_{F_{\mathbf{X}}(\mathbf{x}):\|\mathbf{X}\|\stackrel{a.s.}=R}h(\mathbf{Y})-\ln (2\pi e)
\end{equation}
where $R$ denotes the constant envelope and the capaciy is in $\frac{\mbox{nats}}{\mbox{channel use}}$. The abbreviation $a.s.$ stands for almost surely\footnote{More precisely, Let $\Omega$ be the sample space of the probability model over which the random vector $\mathbf X$ is defined. $\|\mathbf{X}\|\stackrel{a.s.}=R$ is equivalent to $\mbox{Pr}\{\omega\in\Omega|\ \|\mathbf{X}(\omega)\|=R\}=1.$} and $F_{\mathbf X}(\mathbf x)$ denotes the CDF of the input over which the optimization is done.
The pdf of the output determined by the input is given by
\begin{equation}\label{ew}
    f_{\mathbf{Y}}(\mathbf y;F_{\mathbf{X}})=\int\!\!\!\int_{\|\mathbf x\|=R}\frac{1}{2\pi}e^{-\frac{(y_1-\lambda x_1)^2}{2}-\frac{(y_2-x_2)^2}{2}}d^2F_{\mathbf X}(\mathbf x)
\end{equation}
where the notation $;F_{\mathbf{X}}$ is to emphasize that $\mathbf{Y}$ has been resulted by $F_{\mathbf{X}}$. Due to the symmetry of noise, it suffices to consider the input distributions that satisfy the following
\begin{align}
    d^2F_{\mathbf X}(\mathbf x)&=(dF_{X_1}(x_1)).\left[\frac{1}{2}\delta(x_2-\sqrt{R^2-x_1^2})+\frac{1}{2}\delta(x_2+\sqrt{R^2-x_1^2})\right]dx_2\label{op1}
   % f_{\mathbf X}(\mathbf x)&=f_{X_2}(x_2)f_{X_1|X_2}(x_1|x_2)=f_{X_2}(x_2)\left[\frac{1}{2}\delta(x_1-\sqrt{R^2-x_2^2})+\frac{1}{2}\delta(x_1+\sqrt{R^2-x_2^2})\right]\label{op2}.
\end{align}
where $\delta(.)$ is the Dirac-delta function. In other words, any other input distribution that cannot be written as in (\ref{op1}), cannot be an optimal distribution and hence is excluded from our consideration.

Substituting (\ref{op1}) in (\ref{ew}), we get the output pdf as
\begin{align}
    f_{\mathbf{Y}}(\mathbf y;F_{X_1})&=\int_{-R}^{R}K(y_1,y_2,x)dF_{X_1}(x)\label{op3}
    %f_{\mathbf{Y}}(\mathbf y;f_{X_2})&=\int_{-R}^{R}K_2(y_1,y_2,x)f_{X_2}(x)dx\label{op4}
\end{align}
where the kernel is
\begin{align}
    K(y_1,y_2,x)&=\frac{1}{2\pi}e^{-\frac{(y_1-\lambda x)^2}{2}}\left[\frac{1}{2}e^{-\frac{(y_2-\sqrt{R^2-x^2})^2}{2}} + \frac{1}{2}e^{-\frac{(y_2+\sqrt{R^2-x^2})^2}{2}}\right].\label{op5}
    %K_2(y_1,y_2,x)&=\frac{1}{2\pi}e^{-\frac{(y_2- x)^2}{2}}\left[\frac{1}{2}e^{-\frac{(y_1-\lambda\sqrt{R^2-x^2})^2}{2}} + \frac{1}{2}e^{-\frac{(y_1+\lambda\sqrt{R^2-x^2})^2}{2}}\right]\label{op6}.
\end{align}
The marginal entropy density of the output variables induced by the input is defined as \cite{Smith}
\begin{align}
    \tilde{h}_{\mathbf Y}(x;F_{X_1})&=-\int_{-\infty}^{\infty}\!\!\int_{-\infty}^{\infty}K(y_1,y_2,x)\ln f_{\mathbf Y}(\mathbf y;F_{X_1})d\mathbf y
    %\tilde{h}_{\mathbf Y}(x;f_{X_2})&=-\int_{-\infty}^{\infty}\!\!\int_{-\infty}^{\infty}K_2(y_1,y_2,x)\ln f_{\mathbf Y}(\mathbf y;f_{X_2})d\mathbf y
\end{align}
which satisfies the following (which in turn justifies why it is named density)
\begin{align}
    h(\mathbf Y;F_{X_1})&=\int_{-R}^{R}\tilde{h}_{\mathbf Y}(x;F_{X_1})dF_{X_1}(x).
   %  h(\mathbf Y;f_{X_2})&=\int_{-R}^{R}\tilde{h}_{\mathbf Y}(x;f_{X_2})f_{X_2}(x)dx.
\end{align}
Finally, the optimization problem in (\ref{4e2}) becomes equivalent to
\begin{equation}\label{op100}
    C(R) = \sup_{F_{X_1}(x):X_1\in[-R,R]} h(\mathbf{Y};F_{X_1})-\ln (2\pi e)
\end{equation}
where $X_1\in[-R,R]$ is in $a.s.$ sense.
%Assume that we use the representation in (\ref{op1}) for the input distribution.
\section{Main results}\label{th}
Let $\epsilon^*_X$ denote the set of points of increase of the optimal input distribution.\footnote{A point $P$ is said to be a point of increase of a distribution if for any open set $\Gamma$ containing $P$, we have $\mbox{Pr}\{\Gamma\}>0.$}

\textbf{Theorem.} The optimization problem in (\ref{op100}) has a unique solution (denoted by $F^*_{X_1}(x)$) which satisfies the following necessary and sufficient conditions
\begin{align}
    \tilde{h}_{\mathbf Y}(x;F^*_{X_1})&=h(\mathbf Y;F^*_{X_1})\ \ \ \forall x \in \epsilon^*_X\label{op12.2}\\
    \tilde{h}_{\mathbf Y}(x;F^*_{X_1})&<h(\mathbf Y;F^*_{X_1})\ \ \ \forall x \in [-R,R]-\epsilon^*_X\label{op12.3}.
\end{align}
Further, $\epsilon^*_X$ consists of a finite number of mass points in the interval $[-R,R]$ (i.e., $|\epsilon^*_X|<\infty$).
\section{Proof of the theorem}\label{pth}
The steps of the proof are as follows. The uniqueness of the solution along with the necessary and sufficient conditions are obtained through the convex optimization problem. The finite cardinality of $\epsilon^*_X$ is proved by contradiction. In other word, it is shown that infinite number of mass points for the optimal input is not possible.

Let $\mathbb{F}_R$ denote the set of all cumulative distribution functions having their support in the interval $[-R,R]$, i.e.
\begin{equation}\label{op11}
    \mathbb{F}_{R}=\{F_{X_1}(x)|F_{X_1}(x)=0 \ \forall x<-R\ ,\ F_{X_1}(x)=1\  \forall x\geq R\}.
\end{equation}

\textbf{Proposition 1.} The metric space $(\mathbb{F}_{R},d_L)$ is convex and compact where $d_L$ denotes the Levy metric \cite{Loeve}.
\begin{proof}
The proof is the same as that in \cite{Smith2} and \cite[Appendix I]{Abou}.
\end{proof}
\textbf{Proposition 2.} The differential entropy $h(\mathbf Y;F_{X_1}):\mathbb{F}_{R}\to \mathbb{R}$ is continuous.
\begin{proof}
The proof is the same as that in \cite{Smith2}, \cite[Proposition 3]{Shamai}, \cite[Appendix I]{Abou} and \cite[Proposition 1]{Tchamkerten}.
\end{proof}
\textbf{Proposition 3.} The differential entropy $h(\mathbf Y;F_{X_1}):\mathbb{F}_{R}\to \mathbb{R}$ is strictly concave and weakly differentiable.
\begin{proof}
The proof is the same as that in \cite{Smith2}, \cite[Proposition 4]{Shamai}, \cite[Appendix II]{Abou} and \cite[Proposition 2]{Tchamkerten}.
\end{proof}
The weak derivative of $h(\mathbf Y;F_{X_1})$ at $F^0_{X_1}$ is given by
\begin{align}
    h'_{F^0_{X_1}}(\mathbf Y;F_{X_1})&=\lim_{\zeta\to0}\frac{h(\mathbf Y;(1-\zeta)F^0_{X_1}+\zeta F_{X_1})-h(\mathbf Y;F^0_{X_1})}{\zeta}\nonumber\\
    &=   \lim_{\zeta\to0}\frac{\int_{-R}^{R}\tilde{h}_{\mathbf Y}\left(x;(1-\zeta)F^0_{X_1}+\zeta F_{X_1}\right)d\left((1-\zeta)F^0_{X_1}(x)+\zeta F_{X_1}(x)\right)-\int_{-R}^{R}\tilde{h}_{\mathbf Y}\left(x;F^0_{X_1}\right)dF^0_{X_1}(x)}{\zeta}  \nonumber\\
    &= \lim_{\zeta\to0}\frac{(1-\zeta)\int_{-R}^{R}\tilde{h}_{\mathbf Y}(x;F^0_{X_1})dF^0_{X_1}(x)+\zeta\int_{-R}^{R}\tilde{h}_{\mathbf Y}(x;F^0_{X_1})dF_{X_1}(x)-\int_{-R}^{R}\tilde{h}_{\mathbf Y}\left(x;F^0_{X_1}\right)dF^0_{X_1}(x)}{\zeta}  \nonumber\\
    &=\int_{-R}^{R}\tilde{h}_{\mathbf Y}(x;F^0_{X_1})dF_{X_1}(x)-h(\mathbf Y;F^0_{X_1})\ \ ,\ \ \forall F_{X_1}\in\mathbb{F}_{R}.\label{op12}
\end{align}
Since $h(\mathbf Y;F_{X_1})$ is a concave map from $\mathbb{F}_{R}$ to $\mathbb{R}$, Lagrangian optimization \cite{Luenberger} guarantees a unique solution for (\ref{op100}) and the necessary and sufficient condition for the maximizer $F^*_{X_1}$ is
\begin{equation}\label{op13}
    \int_{-R}^{R}\tilde{h}_{\mathbf Y}(x;F^*_{X_1})dF_{X_1}(x)\leq h(\mathbf Y;F^*_{X_1})\ \ ,\ \ \forall F_{X_1}\in\mathbb{F}_{R}.
\end{equation}
It can be shown that (\ref{op13}) is equivalent to (\ref{op12.2}) and (\ref{op12.3}) (as in \cite[Corollary 1]{Smith}). The marginal entropy density can be extended to the complex domain, i.e.
\begin{equation}\label{op14}
    \tilde{h}_{\mathbf Y}(z;F_{X_1})=-\int_{-\infty}^{\infty}\!\!\int_{-\infty}^{\infty}K(y_1,y_2,z)\ln f_{\mathbf Y}(\mathbf y;F_{X_1})d\mathbf y \ \ ,\ \ z\in\mathbb{C}.
\end{equation}
Let $\mathbb{D}=\mathbb{C}-\{(-\infty,-R]\cup[R,+\infty)\}$.

\textbf{Proposition 4}. The kernel $K(y_1,y_2,z)$ is holomorphic on $\mathbb{D}$.
\begin{proof}
This can be verified\footnote{Alternatively, it can be verified by noting that $\mathbb{D}$ is the domain where $\log(R^2-z^2)$ is holomorphic.} by the fact that the real and imaginary parts of $K(y_1,y_2,z=x+jy)$ have continuous partial derivatives and satisfy the Cauchy-Riemann equations on $\mathbb{D}$. As a result, by Cauchy's theorem, for every rectifiable closed curve $\gamma$ in $\mathbb{D}$,
\begin{equation}\label{e36.01}
    \int_{\gamma}K(y_1,y_2,z)dz = 0.
\end{equation}\qedhere
\end{proof}

\textbf{Proposition 5}. The marginal entropy density $\tilde{h}_{\mathbf Y}(z;F_{X_1})$ is holomorphic on $\mathbb{D}$.
\begin{proof}
First, we show the continuity of $\tilde{h}_{\mathbf Y}(z;F_{X_1})$ on $\mathbb{D}$. Let $\{z_m\}_1^{\infty}$ be a sequence of complex numbers in $\mathbb{D}$ converging to $z_0\in\mathbb{D}$. Since $K(y_1,y_2,z)$ is holomorphic on this domain, it is continuous. Therefore,
\begin{equation}\label{e36.001}
    \lim_{m\to\infty}K(y_1,y_2,z_m)\ln f_{\mathbf Y}(\mathbf y;F_{X_1})=K(y_1,y_2,z_0)\ln f_{\mathbf Y}(\mathbf y;F_{X_1}).
\end{equation}
By the application of Lebesgue's dominated convergence theorem, the continuity and boundedness of the kernel guarantees the continuity of $f_{\mathbf Y}(\mathbf y;F_{X_1})$ given in (\ref{op3}). %(See Lemma 3 in \cite{Tchamkerten}).
This allows us to write
\begin{equation}\label{e36.1}
   \min_{x\in[-R,R]}K(y_1,y_2,x)\leq f_{\mathbf Y}(\mathbf y;F_{X_1})\leq\max_{x\in[-R,R]}K(y_1,y_2,x).
\end{equation}
Therefore,
\begin{align}\label{bound}
    \frac{1}{2\pi}e^{-\frac{y_1^2+y_2^2}{2}-\frac{\lambda^2R^2}{2}-\lambda R|y_1|}\leq f_{\mathbf Y}(\mathbf y;F_{X_1})\leq \frac{1}{2\pi}e^{-\frac{y_1^2+y_2^2}{2}-\frac{R^2}{2}+\lambda R|y_1|}\cosh Ry_2
\end{align}
which results in
\begin{equation}\label{opup}
    |\ln f_{\mathbf Y}(\mathbf y;F_{X_1})|\leq \ln(2\pi) + \frac{y_1^2+y_2^2}{2}+\frac{\lambda^2R^2}{2}+\lambda R|y_1|+\ln(\cosh Ry_2).
\end{equation}
It can be verified that
\begin{align}
    |\tilde{h}_{\mathbf Y}(z_m;F_{X_1})|&\leq\int_{-\infty}^{\infty}\!\!\int_{-\infty}^{\infty}|K(y_1,y_2,z_m)||\ln f_{\mathbf Y}(\mathbf y;F_{X_1})|d\mathbf y\nonumber\\
    &\leq|\frac{1}{2\pi}e^{-\frac{(\lambda^2-1)z_m^2}{2}}|e^{-\frac{R^2}{2}}\int_{-\infty}^{\infty}\!\!\int_{-\infty}^{\infty}e^{-\frac{y_1^2+y_2^2}{2}}|e^{\lambda y_1z_m}||\cosh(y_2\sqrt{R^2-z_m^2})||\ln f_{\mathbf Y}(\mathbf y;F_{X_1})|d\mathbf y\nonumber\\
    &\leq |\frac{1}{2\pi}e^{-\frac{(\lambda^2-1)z_m^2}{2}}|e^{-\frac{R^2}{2}}\int_{-\infty}^{\infty}\!\!\int_{-\infty}^{\infty}e^{-\frac{y_1^2+y_2^2}{2}}e^{\lambda y_1\mbox{Re}(z_m)}e^{|y_2|\sqrt{R^2+|z_m|^2}}|\ln f_{\mathbf Y}(\mathbf y;F_{X_1})|d\mathbf y\label{op1.1}\\
    &<\infty \label{op1.2}
\end{align}
where in (\ref{op1.1}), we have used the fact that $|e^z|=e^{\mbox{Re}(z)}$, $|\cosh(z)|\leq\cosh(\mbox{Re}(z))$ and $\cosh(x)\leq e^{|x|} (x\in\mathbb{R})$. (\ref{op1.2}) is due to the upper bound in (\ref{opup}) and the term $e^{-\frac{y_1^2+y_2^2}{2}}$ in the integration. Since the absolute value of the integrand of $\tilde{h}_{\mathbf Y}(z_m;F_{X_1})$ is integrable, by Lebesgue's dominated convergence theorem, we have
\begin{align}
    \lim_{m\to\infty}\tilde{h}_{\mathbf Y}(z_m;F_{X_1})&=-\lim_{m\to\infty}\int_{-\infty}^{\infty}\!\!\int_{-\infty}^{\infty}K(y_1,y_2,z_m)\ln f_{\mathbf Y}(\mathbf y;F_{X_1})d\mathbf y\nonumber\\
    &=-\int_{-\infty}^{\infty}\!\!\int_{-\infty}^{\infty}\lim_{m\to\infty}K(y_1,y_2,z_m)\ln f_{\mathbf Y}(\mathbf y;F_{X_1})d\mathbf y\nonumber\\
   &=-\int_{-\infty}^{\infty}\!\!\int_{-\infty}^{\infty}K(y_1,y_2,z_0)\ln f_{\mathbf Y}(\mathbf y;F_{X_1})d\mathbf y\nonumber\\
   &=\tilde{h}_{\mathbf Y}(z_0;F_{X_1})
\end{align}
which proves the continuity of $\tilde{h}_{\mathbf Y}(z;F_{X_1})$. Let $\partial T$ denote an arbitrary triangle in $\mathbb{D}$. We can write,
\begin{align}
   \int_{\partial T}\tilde{h}_{\mathbf Y}(z;F_{X_1})dz&=-\int_{\partial T}\int_{-\infty}^{\infty}\!\!\int_{-\infty}^{\infty}K(y_1,y_2,z)\ln f_{\mathbf Y}(\mathbf y;F_{X_1})d\mathbf ydz\nonumber\\
   &=-\int_{-\infty}^{\infty}\!\!\int_{-\infty}^{\infty}\int_{\partial T}K(y_1,y_2,z)dz\ln f_{\mathbf Y}(\mathbf y;F_{X_1})d\mathbf y\label{e36.02}\\
   &=0\label{e36.03}
\end{align}
where (\ref{e36.02}) is allowed by Fubini's theorem, because for a given rectifiable triangle $\partial T$,
\begin{equation}\label{e36.003}
    \int_{\partial T}|\tilde{h}_{\mathbf Y}(z;F_{X_1})|dz<\infty.
\end{equation}
(\ref{e36.03}) is due to the holomorphy of $K(y_1,y_2,z)$ (see (\ref{e36.01})). Therefore, by Morera's theorem (with weakened hypothesis), it is concluded that $\tilde{h}_{\mathbf Y}(z;F_{X_1})$ is holomorphic on $\mathbb{D}$.
\end{proof}
If $\epsilon^*_X$ has an infinite number of points, since it is bounded in $[-R,R]$, it must have an accumulation point by Bolzano-Weierstrass theorem. If the accumulation point is in $(-R,R)$, it is also in the domain where the marginal entropy density is holomorphic (i.e., $\mathbb{D}=\mathbb{C}-\{(-\infty,-R]\cup[R,+\infty)\}$). Therefore, by using the identity theorem of holomorphic functions of one complex variable, the following must be satisfied
\begin{equation}\label{op7}
  \tilde{h}_{\mathbf Y}(z;F^*_{X_1})=h(\mathbf Y;F^*_{X_1})\ \ ,\ \ \forall z \in \mathbb{D}.
\end{equation}
If the accumulation point is on the boundary (i.e. it is $\pm R$) where the holomorphy fails to hold (and the usage of identity theorem is not allowed), we can still show that (\ref{op7}) must hold. The reason is as follows. Note that an accumulation point of $P$ ($\in [-R,R]$) on the $x_1$ axis is equivalent to an accumulation point of $\sqrt{R^2-P^2}$ ($\in [-R,R]$) on the $x_2$ axis and vice versa. Therefore, if there is an accumulation point of $\pm R$ on $x_1$ axis, there is an accumulation point of $0$ on $x_2$ axis. By using an alternative representation of the input distribution in (\ref{op1}), we can write
\begin{equation}\label{alt}
    d^2F_{\mathbf X}(\mathbf x)=(dF_{X_2}(x_2)).\left[\frac{1}{2}\delta(x_1-\sqrt{R^2-x_2^2})+\frac{1}{2}\delta(x_1+\sqrt{R^2-x_2^2})\right]dx_1
\end{equation}
which results in an equivalent optimization problem as
\begin{equation}\label{op101}
    C(R) = \sup_{F_{X_2}(x):X_2\in[-R,R]} h(\mathbf{Y};F_{X_2})-\ln (2\pi e)
\end{equation}
with the following modified terms
\begin{align}
    f_{\mathbf{Y}}(\mathbf y;F_{X_2})&=\int_{-R}^{R}K'(y_1,y_2,x)dF_{X_2}(x)\\
    \tilde{h}_{\mathbf Y}(x;F_{X_2})&=-\int_{-\infty}^{\infty}\!\!\int_{-\infty}^{\infty}K'(y_1,y_2,x)\ln f_{\mathbf Y}(\mathbf y;F_{X_2})d\mathbf y\\
    K'(y_1,y_2,x)&=\frac{1}{2\pi}e^{-\frac{(y_2- x)^2}{2}}\left[\frac{1}{2}e^{-\frac{(y_1-\lambda\sqrt{R^2-x^2})^2}{2}} + \frac{1}{2}e^{-\frac{(y_1+\lambda\sqrt{R^2-x^2})^2}{2}}\right].
\end{align}
By using the same tools in analysis, an accumulation point of $\pm R$ on $x_1$ axis (which is equivalent to an accumulation point of $0$ on $x_2$ axis) results in
\begin{equation}\label{op8}
        \tilde{h}_{\mathbf Y}(z;F^*_{X_2})=h(\mathbf Y;F^*_{X_2})\ \ ,\ \ \forall z \in \mathbb{D}.
\end{equation}
This also means that all the points on the $x_2$ axis in the interval $(-R,R)$ are points of increase of $F^*_{X_2}$. Hence, all the points on the $x_1$ axis in the interval $(-R,R)-\{0\}$ are points of increase of $F^*_{X_1}$ which in turn results in having an accumulation point ($\neq\pm R$) on the $x_1$ axis. Therefore, regardless of having the accumulation point on $x_1$ axis in the interior or the boundary of $[-R,R]$, the assumption of having an infinite number of mass points in $\epsilon^*_X$ results in (\ref{op7}). In what follows, the equality in (\ref{op7}) is disproved\footnote{Note that this is the equality mentioned in section \ref{Intro}.}. Rewriting (\ref{op7}), we have
\begin{align}
  -\frac{1}{2\pi}\int_{-\infty}^{\infty}\!\!\int_{-\infty}^{\infty}e^{-\frac{(y_1-\lambda z)^2}{2}}\left[\frac{1}{2}e^{-\frac{(y_2-\sqrt{R^2-z^2})^2}{2}} + \frac{1}{2}e^{-\frac{(y_2+\sqrt{R^2-z^2})^2}{2}}\right]\ln f_{\mathbf Y}(\mathbf y;F^*_{X_1})d\mathbf y=c\ \  ,\ \ \forall z \in \mathbb{D}\label{op9}
\end{align}
where $c(=h(\mathbf Y;F^*_{X_1}))$ is a constant. Let $z=x+i\epsilon\ (\epsilon\neq 0)$. For any given $x<\infty$, (\ref{op9}) implies
\begin{align}
    \lim_{\epsilon\to 0}-\frac{1}{2\pi}\int_{-\infty}^{\infty}\!\!\int_{-\infty}^{\infty}e^{-\frac{(y_1-\lambda (x+i\epsilon))^2}{2}}\left[\frac{1}{2}e^{-\frac{(y_2-\sqrt{R^2-(x+i\epsilon)^2})^2}{2}} + \frac{1}{2}e^{-\frac{(y_2+\sqrt{R^2-(x+i\epsilon)^2})^2}{2}}\right]\ln f_{\mathbf Y}(\mathbf y;F^*_{X_1})d\mathbf y=c.\label{abs}
\end{align}
Since the absolute value of the integrand in (\ref{abs}) is integrable, by the application of Lebesgue's dominated convergence theorem, we can take the limit inside the integrals and obtain
\begin{equation}\label{op10}
    -\frac{1}{2\pi}\int_{-\infty}^{\infty}\!\!\int_{-\infty}^{\infty}e^{-\frac{(y_1-\lambda x)^2}{2}}\left[\frac{1}{2}e^{-\frac{(y_2-\sqrt{R^2-x^2})^2}{2}} + \frac{1}{2}e^{-\frac{(y_2+\sqrt{R^2-x^2})^2}{2}}\right]\ln f_{\mathbf Y}(\mathbf y;F^*_{X_1})d\mathbf y=c\ \ ,\ \ \forall x\in\mathbb{R}.
\end{equation}
In the sequel, it is shown that (\ref{op10}) does not hold. More precisely, it is shown that the left hand side of (\ref{op10}) becomes unbounded as $x$ goes to infinity and therefore it cannot be a constant on the whole real line. We rewrite $f_{\mathbf Y}(\mathbf y;F^*_{X_1})$ as
\begin{equation}\label{at1}
    f_{\mathbf Y}(\mathbf y;F^*_{X_1})=\frac{1}{2\pi}e^{-\frac{y_1^2+y_2^2}{2}-\frac{R^2}{2}}g(\mathbf y;F^*_{X_1})
\end{equation}
where
\begin{equation}\label{at2}
    g(\mathbf y;F^*_{X_1})=\int_{-R}^{R}e^{-\frac{(\lambda^2-1)x^2}{2}+\lambda y_1x}\cosh(y_2\sqrt{R^2-x^2})dF^*_{X_1}(x).
\end{equation}
By substituting (\ref{at1}) in (\ref{op10}), we obtain
\begin{equation}\label{at3}
    \ln(2\pi)+ R^2 + 1 + \frac{\lambda^2-1}{2}x^2-\underbrace{\int_{-\infty}^{\infty}\!\!\int_{-\infty}^{\infty}K(y_1,y_2,x)\ln g(\mathbf y;F^*_{X_1})dy_2dy_1}_{I}=c\ \ ,\ \ \forall x\in\mathbb{R}.
\end{equation}
The double integral in (\ref{at3}) at large values of $x$ can be written as (note that we make use of the equality $\cosh(ix)=\cos(x)$)
\begin{align}
    I&=e^{-\frac{(\lambda^2-1)}{2}x^2}\int_{-\infty}^{\infty}\!\!\int_{-\infty}^{\infty}e^{\frac{-y_1^2-y_2^2+2\lambda y_1x}{2}}\cos (y_2\sqrt{x^2-R^2})\ln g(\mathbf y;F^*_{X_1})dy_2dy_1\\
    &=\lim_{a,b,c,d\to+\infty}e^{-\frac{(\lambda^2-1)}{2}x^2}\int_{-a}^{b}\!\!\int_{-c}^{d}e^{\frac{-y_1^2-y_2^2+2\lambda y_1x}{2}}\cos (y_2\sqrt{x^2-R^2})\ln g(\mathbf y;F^*_{X_1})dy_2dy_1\label{def}.
\end{align}
If it can be shown that $|I|\leq O(x)$ (i.e., the growth of $I$ with $x$ is at most linearly), then the proof is complete by observing that the left hand side of (\ref{at3}) does not converge to any real number as $x$ increases and therefore it cannot be a constant on the whole real line.

Let $M,K$ be two sufficiently large numbers satisfying $K\gg M$ and define $I(M)$ as
%\begin{align}
%    I(M)&= e^{-\frac{(\lambda^2-1)}{2}x^2}\int_{-\infty}^{\infty}\!\!\int_{-M}^{M}e^{\frac{-y_1^2-y_2^2+2\lambda y_1x}{2}}\cos (y_2\sqrt{x^2-R^2})\ln g(\mathbf y;F^*_{X_1})dy_2dy_1\label{at4}
%\end{align}
%Therefore, we can write
%\begin{equation}\label{had}
%    I = \lim_{M\to+\infty} I(M).
%\end{equation}
%&=I_{-\infty}^{-K} + I_{-K}^{K} + I_{K}^{+\infty}\label{at5}
%where (\ref{at4}) is due to the term $e^{-\frac{y_2^2}{2}}$ in the integration\footnote{It is clear that the same truncation of the boundary of integration cannot be applied to $y_1$ due to the term $e^{\lambda y_1x}$ when x goes to infinity.} and $0<M<\infty$ is sufficiently large.
\begin{equation}\label{at5}
    I(M) = I_{-\infty}^{-K} + I_{-K}^{K} + I_{K}^{+\infty}
\end{equation}
in which
\begin{align}
    I_{a}^{b} \triangleq e^{-\frac{(\lambda^2-1)}{2}x^2}\int_{a}^{b}\!\!\int_{-M}^{M}e^{\frac{-y_1^2-y_2^2+2\lambda y_1x}{2}}\cos (y_2\sqrt{x^2-R^2})\ln g(\mathbf y;F^*_{X_1})dy_2dy_1\ \ a,b\in \mathbb{R}\cup\{-\infty,+\infty\}.
\end{align}
In what follows, we find upper bounds for each of the terms in (\ref{at5}) when $x$ is sufficiently large.
\begin{align}
    \lim_{x\to+\infty}|I_{-K}^{K}|&\leq\lim_{x\to+\infty}e^{-\frac{(\lambda^2-1)}{2}x^2}\int_{-K}^{K}\!\!\int_{-M}^{M}e^{\frac{-y_1^2-y_2^2+2\lambda y_1x}{2}}|\cos (y_2\sqrt{x^2-R^2})||\ln g(\mathbf y;F^*_{X_1})|dy_2dy_1\nonumber\\
    &\leq\lim_{x\to+\infty}e^{-\frac{(\lambda^2-1)}{2}x^2}\int_{-K}^{K}\!\!\int_{-M}^{M}e^{\frac{-y_1^2-y_2^2+2\lambda y_1x}{2}}|\ln g(\mathbf y;F^*_{X_1})|dy_2dy_1\nonumber\\
    &\leq\lim_{x\to+\infty}e^{-\frac{(\lambda^2-1)}{2}x^2}\int_{-K}^{K}\!\!\int_{-M}^{M}e^{\frac{-y_1^2-y_2^2+2\lambda y_1x}{2}}R(\lambda|y_1|+|y_2|)dy_2dy_1\label{chera}\\
    &=0
\end{align}
where in (\ref{chera}), we use the following upper bound for $g(\mathbf y;F^*_{X_1})$ defined in (\ref{at2})
\begin{equation}\label{ezaf}
     g(\mathbf y;F^*_{X_1})\leq e^{R(\lambda|y_1|+|y_2|)}.
\end{equation}
Similarly, for the term $I_{-\infty}^{-K}$, we can write
\begin{align}
    \lim_{x\to+\infty}|I_{-\infty}^{-K}|&\leq\lim_{x\to+\infty}e^{-\frac{(\lambda^2-1)}{2}x^2}\int_{-\infty}^{-K}\!\!\int_{-M}^{M}e^{\frac{-y_1^2-y_2^2+2\lambda y_1x}{2}}R(\lambda|y_1|+|y_2|)dy_2dy_1\\
    &=0.
\end{align}
Bounding $I_{K}^{+\infty}$ is more involved. First, according to the boundary of integration in $I_{K}^{+\infty}$, we have $y_1\geq K\gg M\geq |y_2|$. By rewriting $g(\mathbf y;F^*_{X_1})$ in this regime, we get
\begin{align}
    g(\mathbf y;F^*_{X_1})&=\int_{-R}^{R}e^{-\frac{(\lambda^2-1)x^2}{2}+\lambda y_1x}\cosh(y_2\sqrt{R^2-x^2})dF^*_{X_1}(x)\nonumber
\end{align}
\begin{align}
    &=\int_{-R}^{R}\left [\frac{1}{2}e^{-\frac{(\lambda^2-1)x^2}{2}+\lambda y_1x+y_2\sqrt{R^2-x^2}}+\frac{1}{2}e^{-\frac{(\lambda^2-1)x^2}{2}+\lambda y_1x-y_2\sqrt{R^2-x^2}}\right ]dF^*_{X_1}(x)\nonumber\\
    &\approx \int_{-R}^{R}\left [\frac{1}{2}e^{\lambda y_1x}+\frac{1}{2}e^{\lambda y_1x}\right ]dF^*_{X_1}(x) \label{at6}\\
    &=\int_{-R}^{R}e^{\lambda y_1x}dF^*_{X_1}(x)
\end{align}
where (\ref{at6}) is due to the fact that $y_1\geq K\gg M\geq \max\{|y_2|,R\}$ and this approximation gets better when $M\to\infty$ and $K$ grows faster than $M$. Therefore,
\begin{align}
    &\ \ \ \ \lim_{M\to+\infty : x\gg K\gg M}|I_{K}^{+\infty}|\\
    &=\lim_{M\to+\infty : x\gg K\gg M}\left|e^{-\frac{(\lambda^2-1)}{2}x^2}\int_{K}^{\infty}\!\!\int_{-M}^{M}e^{\frac{-y_1^2-y_2^2+2\lambda y_1x}{2}}\cos (y_2\sqrt{x^2-R^2})\ln g(\mathbf y;F^*_{X_1})dy_2dy_1\right|\nonumber\\
    &\approx\lim_{M\to+\infty : x\gg K\gg M}\left|e^{-\frac{(\lambda^2-1)}{2}x^2}\int_{K}^{\infty}e^{\frac{-y_1^2+2\lambda y_1x}{2}}\!\!\int_{-M}^{M}e^{-\frac{y_2^2}{2}}\cos (y_2\sqrt{x^2-R^2})dy_2\ln\left(\int_{-R}^{R}e^{\lambda y_1x}dF^*_{X_1}(x)\right) dy_1\right|\nonumber\\
    &=\lim_{K\to+\infty : x\gg K}\left|\sqrt{2\pi}e^{-\frac{\lambda^2}{2}x^2+\frac{R^2}{2}}\int_{K}^{\infty}e^{\frac{-y_1^2+2\lambda y_1x}{2}}\ln\left(\int_{-R}^{R}e^{\lambda y_1x}dF^*_{X_1}(x)\right) dy_1\right|\label{at7}\\
    &\leq \lim_{K\to+\infty : x\gg K}\sqrt{2\pi}e^{-\frac{\lambda^2}{2}x^2+\frac{R^2}{2}}\int_{K}^{\infty}e^{\frac{-y_1^2+2\lambda y_1x}{2}}R\lambda y_1dy_1\nonumber\\
    &=\lim_{K\to+\infty : x\gg K}\sqrt{2\pi}R\lambda e^{\frac{R^2}{2}}\int_{K}^{\infty}e^{-\frac{(y_1-\lambda x)^2}{2}}y_1dy_1\nonumber\\
    &=\lim_{K\to+\infty : x\gg K}\sqrt{2\pi}R\lambda e^{\frac{R^2}{2}}\left(e^{-\frac{(K-\lambda x)^2}{2}}+\sqrt{2\pi}\lambda xQ(K-\lambda x)\right)\label{at8}\\
    &=\lim_{x\to+\infty}2\pi Re^{\frac{R^2}{2}}\lambda^2x
\end{align}
where in (\ref{at7}), we have used the equality $\int_{-\infty}^{+\infty}e^{-\beta x^2}\cos bx dx=\sqrt{\frac{\pi}{\beta}}e^{-\frac{b^2}{4\beta}}\ (\mbox{Re}\{\beta\}>0)$ and this approximation becomes better as $M$ grows. In (\ref{at8}), $Q(a)=\int_{a}^{\infty}\frac{e^{-\frac{t^2}{2}}}{\sqrt{2\pi}}dt$. The limit of the left hand side of (\ref{at3}) is
\begin{align}
    \lim_{x\to+\infty}\ln(2\pi)+ R^2 + 1 + \frac{\lambda^2-1}{2}x^2-I&= \lim_{M\to+\infty : x\gg K\gg M}\ln(2\pi)+ R^2 + 1 + \frac{\lambda^2-1}{2}x^2-I(M)\\
    &=\lim_{M\to+\infty : x\gg K\gg M}\ln(2\pi)+ R^2 + 1 + \frac{\lambda^2-1}{2}x^2-I_{-\infty}^{-K}-I_{-K}^{K}-I_{K}^{+\infty}\nonumber\\
    &\geq\lim_{M\to+\infty : x\gg K\gg M}\ln(2\pi)+ R^2 + 1 + \frac{\lambda^2-1}{2}x^2-|I_{-\infty}^{-K}|-|I_{-K}^{K}|-|I_{K}^{+\infty}|\nonumber\\
    &\geq\lim_{x\to+\infty}\ln(2\pi)+ R^2 + 1 + \frac{\lambda^2-1}{2}x^2-2\pi Re^{\frac{R^2}{2}}\lambda^2x\nonumber\\
    &=+\infty.\label{inft}
\end{align}
Note that the assumption of $\lambda>1$ is crucial for all of the bounds specially in (\ref{inft}). Therefore, (\ref{at3}) does not hold on the whole real line (and in turn (\ref{op7}) does not hold on $\mathbb{D}$) which makes the assumption of infinite number of mass points incorrect. This completes the proof.
\section{Asymptotic behavior}\label{asymbeh}
\textbf{Corollary 1}. For $\lambda\geq 1$, when the norm of the input vector is very small, we have
\begin{equation}\label{asy1}
  C(R)\approx\frac{\lambda^2R^2}{2}\ \ ,\ \ R\to 0
\end{equation}
and the asymptotically optimal input distribution is given by
\begin{equation}\label{asy2}
    F^{\mbox{asym}}_{\mathbf X}(\mathbf x)=\left[\frac{1}{2}u(x_1-R)+\frac{1}{2}u(x_1+R)\right]u(x_2)
\end{equation}
where $u(.)$ is the unit step function.
\begin{proof}
From (\ref{4e2}), we can write
\begin{align}
    C(R)&\leq \sup_{F_{\mathbf X}(\mathbf x): E[\|X\|^2]\leq R^2} I(\mathbf X;\mathbf Y)\nonumber\\
    &=\frac{1}{2}\ln(1+\lambda^2P^*_1)+\frac{1}{2}\ln(1+P^*_2)\label{as2.2}
\end{align}
where the solutions of the water filling algorithm are given by
\begin{equation}\label{as2}
(P^*_1,P^*_2)=\left\{\begin{array}{cc} (R^2,0) & R^2\leq1-\frac{1}{\lambda^2}\\ (\frac{R^2+1-\frac{1}{\lambda^2}}{2},\frac{R^2-1+\frac{1}{\lambda^2}}{2}) & \mbox{o.w.} \end{array}\right..
\end{equation}
When $R$ is vanishingly small, from (\ref{as2.2}) and (\ref{as2}), we have
\begin{equation}\label{as3}
    \lim_{R\to 0}C(R)\leq\lim_{R\to 0}\frac{1}{2}\ln(1+\lambda^2R^2)=\frac{\lambda^2R^2}{2}.
\end{equation}
In what follows, we show that the distribution in (\ref{asy2}) achieves the upper bound in (\ref{as3}) asymptotically. The pdf of the output induced by this input distribution is
\begin{equation}\label{as4}
    f_{\mathbf Y}(\mathbf y;F^{\mbox{asym}}_{\mathbf X})=\frac{1}{2\pi}e^{-\frac{y_2^2}{2}}\left(\frac{1}{2}e^{-\frac{(y_1-\lambda R)^2}{2}}+\frac{1}{2}e^{-\frac{(y_1+\lambda R)^2}{2}}\right).
\end{equation}
When $R$ is small,
\begin{align}
    h(\mathbf Y;F^{\mbox{asym}}_{\mathbf X})&=\lim_{R\to 0}-\int_{-\infty}^{\infty}\!\!\int_{-\infty}^{\infty}f_{\mathbf Y}(\mathbf y;F^{\mbox{asym}}_{\mathbf X})\ln f_{\mathbf Y}(\mathbf y;F^{\mbox{asym}}_{\mathbf X})d\mathbf y\nonumber\\
    &=\lim_{R\to 0}\int_{-\infty}^{\infty}\!\!\int_{-\infty}^{\infty}\frac{1}{2\pi}e^{-\frac{y_2^2}{2}}\left(\frac{1}{2}e^{-\frac{(y_1-\lambda R)^2}{2}}+\frac{1}{2}e^{-\frac{(y_1+\lambda R)^2}{2}}\right)\left(\ln2\pi+\frac{y_2^2}{2}+\frac{y_1^2}{2}+\frac{\lambda^2R^2}{2}-\ln\cosh(\lambda Ry_1)\right)d\mathbf y\nonumber\\
    &=\lim_{R\to 0}\int_{-\infty}^{\infty}\!\!\int_{-\infty}^{\infty}\frac{1}{2\pi}e^{-\frac{y_2^2}{2}}\left(\frac{1}{2}e^{-\frac{(y_1-\lambda R)^2}{2}}+\frac{1}{2}e^{-\frac{(y_1+\lambda R)^2}{2}}\right)\left(\ln2\pi+\frac{y_2^2}{2}+\frac{y_1^2}{2}+\frac{\lambda^2R^2}{2}-\frac{\lambda^2R^2y_1^2}{2}\right)d\mathbf y\label{as6}\\
    &=\lim_{R\to 0} 1+\ln2\pi+\frac{\lambda^2R^2}{2}\label{as7}
\end{align}
where in (\ref{as6}), we have used the approximations $\cosh x\approx1+\frac{x^2}{2}$ and $\ln(1+x)\approx x$ for $x \ll1$ and in (\ref{as7}), we have dropped the higher order terms of $R$. Therefore, when the norm of the input is very small, the mutual information resulted by the input distribution $F^{\mbox{asym}}_{\mathbf X}(\mathbf x)$ is
\begin{equation}
    \lim_{R\to 0} h(\mathbf Y;F^{\mbox{asym}}_{\mathbf X})-\ln2\pi e=\frac{\lambda^2R^2}{2}
\end{equation}
which confirms that the upper bound in (\ref{as3}) is asymptotically tight.

The asymptotic optimality of the distribution in (\ref{asy2}) can alternatively be proved by inspecting the behavior of the marginal entropy density $\tilde{h}_{\mathbf Y}(x;F_{X_1})$ when $R$ is sufficiently small.
From (\ref{bound}), we have
\begin{equation}
    f_{\mathbf Y}(\mathbf y;F_{X_1})\stackrel{R\to 0}\longrightarrow\frac{1}{2\pi}e^{-\frac{y_1^2+y_2^2}{2}}.
\end{equation}
Therefore,
\begin{align}
    \tilde{h}_{\mathbf Y}(x;F_{X_1})&=-\int_{-\infty}^{\infty}\!\!\int_{-\infty}^{\infty}K(y_1,y_2,x)\ln f_{\mathbf Y}(\mathbf y;F_{X_1})d\mathbf y\\
    &\rightarrow\int_{-\infty}^{\infty}\!\!\int_{-\infty}^{\infty}\frac{1}{2\pi}e^{-\frac{(y_1-\lambda x)^2}{2}}\left[\frac{1}{2}e^{-\frac{(y_2-\sqrt{R^2-x^2})^2}{2}} + \frac{1}{2}e^{-\frac{(y_2+\sqrt{R^2-x^2})^2}{2}}\right](\ln2\pi + \frac{y_1^2+y_2^2}{2} )d\mathbf y\\
    &=1+\ln2\pi+\frac{R^2}{2}+\frac{\lambda^2-1}{2}x^2
\end{align}
which is a strictly convex (and even) function. Hence, the necessary and sufficient
conditions in (\ref{op12.2}) and (\ref{op12.3}) are satisfied if and only if the input is distributed as (\ref{asy2})\footnote{It can alternatively be verified that when $R\ll1$, $\tilde{h}_{\mathbf Y}(x;F_{X_2})$ becomes strictly concave and one mass point at zero on the $x_2$ axis is optimal which is equivalent to (\ref{asy2}). }. Note that in contrast to the optimal distribution, the asymptotically optimal distribution is not unique. As a special case, when $\lambda=1$, the distribution in (\ref{asy2}) with two mass points is still asymptotically optimal. However, the optimal input distribution has an infinite number of mass points uniformly distributed on the circle with radius $R$ (as shown in \cite{Wyner}).
\end{proof}
\textbf{Corollary 2}. For high SNR values, we have
\begin{equation}\label{as20}
    \lim_{R\to\infty}\frac{C(R)}{\ln R}=1.
\end{equation}
In other words, the constant envelope signaling in a 2 by 2 channel has only one degree of freedom.
\begin{proof}
By writing the input of the channel in polar coordinates as $\mathbf{X}=R[\cos\Theta\ ,\ \sin\Theta]^T$, the differential entropy of the input in polar coordinates is given by
\begin{align}
    h(\mathbf X)&=-\int_{\|\mathbf x\|=R}f_{\mathbf X}(\mathbf x)\ln f_{\mathbf X}(\mathbf x)d\mathbf x \nonumber\\
    &=-\int_{0}^{2\pi}f_{\mathbf X}(\mathbf x(R, \theta))\ln f_{\mathbf X}(\mathbf x(R, \theta))|\frac{\partial \mathbf x}{\partial(R, \theta)}|d\theta \nonumber\\
    &=-\int_{0}^{2\pi}f_{\Theta}(\theta)\ln \frac{f_{\Theta}(\theta)}{|\frac{\partial \mathbf x}{\partial(R, \theta)}|}d\theta \nonumber\\
    &=h(\Theta) + \ln R\nonumber\\
    &\leq \ln 2\pi R \label{ach}
\end{align}
where $\frac{\partial \mathbf x}{\partial(R, \theta)}=R$ is the Jacobian of the transform and the maximum in (\ref{ach}) is achieved iff the $\Theta\sim U[0,2\pi).$ The capacity is bounded below as follows.
\begin{align}
    C(R) &= \sup_{F_{\mathbf X}(\mathbf x):\|\mathbf X\|=R} h(\mathbf{Y};F_{\mathbf X})-\ln (2\pi e)\nonumber\\
    &\geq \sup_{F_{\mathbf X}(\mathbf x):\|\mathbf X\|=R} \ln\left(e^{h(\mathbf H\mathbf X)}+e^{h(\mathbf W)}\right)-\ln (2\pi e)\label{q1}\\
    &=\sup_{F_{\mathbf X}(\mathbf x):\|\mathbf X\|=R} \ln\left(e^{\ln|\mbox{det}(\mathbf H)| + h(\mathbf X)}+ 2\pi e\right)-\ln (2\pi e)\nonumber\\
    &= \ln(2\pi\lambda R+2\pi e)-\ln (2\pi e)\label{q3}
\end{align}
where (\ref{q1}) is due to the vector entropy-power inequality (EPI) and in (\ref{q3}) the upper bound in (\ref{ach}) is used.

The capacity is bounded above as follows.
\begin{align}
    C(R) &= \sup_{F_{\mathbf X}(\mathbf x):\|\mathbf X\|=R} h(\mathbf H\mathbf X+\mathbf W;F_{\mathbf X})-\ln (2\pi e)\nonumber\\
    &\leq \sup_{F_{\mathbf X}(\mathbf x):\|\mathbf X\|=R} h(\mathbf H\mathbf X,\mathbf W;F_{\mathbf X})-\ln (2\pi e)\nonumber\\
    &\leq \sup_{F_{\mathbf X}(\mathbf x):\|\mathbf X\|=R} h(\mathbf H\mathbf X;F_{\mathbf X})+h(\mathbf W)-\ln (2\pi e)\nonumber\\
    &= \ln(2\pi R)+\ln \lambda\label{q5}.
\end{align}
Combining (\ref{q3}) and (\ref{q5}), we have
\begin{equation}\label{q6}
   \ln(2\pi\lambda R+2\pi e)-\ln (2\pi e) \leq C(R)\leq \ln(2\pi R)+\ln \lambda.
\end{equation}
Dividing by $\ln R$ and letting $R\to\infty$ results in (\ref{as20}).
\end{proof}
The analysis can be readily generalized to the n-dimensional full rank channels by noting that
\begin{equation}
    h(\mathbf X)\leq \ln\left(\frac{2\pi^{\frac{n}{2}}R^{n-1}}{\Gamma(\frac{n}{2})}\right)
\end{equation}
which is tight iff the distribution of the phase vector of $\mathbf X$ in the spherical coordinates is as follows
\begin{equation}\label{q6.99}
    f_{\mathbf \Theta}(\mathbf \theta)=\frac{1}{2\pi}\prod_{i=1}^{n-2}\alpha_i^{-1}\sin^{n-i-1}\theta_i
\end{equation}
where $\alpha_i = \frac{\sqrt{\pi}\Gamma(\frac{n-i}{2})}{\Gamma(\frac{n-i+1}{2})}$.
Therefore, we have
\begin{equation}\label{q7}
   \frac{n}{2}\ln\left((\frac{2R^{n-1}\pi^{\frac{n}{2}}|\mbox{det}(\mathbf H)|}{\Gamma(\frac{n}{2})})^{\frac{2}{n}}+2\pi e\right)-\frac{n}{2}\ln (2\pi e) \leq C(R)\leq \ln\left(\frac{2\pi^{\frac{n}{2}} R^{n-1}}{\Gamma(\frac{n}{2})}\right)+\ln |\mbox{det}(\mathbf H)|
\end{equation}
which results in
\begin{equation}\label{as3}
    \lim_{R\to\infty}\frac{C(R)}{\ln R}=n-1.
\end{equation}
Intuitively, that loss of 1 degree of freedom is due to the fact that for a constant norm n-dimensional vector, given its $n-1$ elements, the remaining element has the uncertainty of at most 1 bit which does not scale with $R$ as it goes to infinity. Finally, note that the phase distribution in (\ref{q6.99}) is equivalent to uniform distribution on the surface of the hypersphere with radius $R$ which is optimal in the DoF sense\footnote{It is important to note that the DoF-achieving distribution is not unique in contrast to the capacity-achieving distribution.}.
%which results in
%\begin{equation}\label{op11}
%    e^{-\frac{(\lambda^2-1)}{2}x^2}\int_{-\infty}^{\infty}\!\!\int_{-\infty}^{\infty}e^{\frac{-y_1^2-y_2^2+2\lambda y_1x}{2}}\cosh (y_2\sqrt{R-x^2})\ln f_{\mathbf Y}(\mathbf y;f^*_{X_1})d\mathbf y=c\ \ \forall x\in\mathbb{R}.
%\end{equation}
%\color{red}There is an integral somehow similar (if we could ignore $R$) to our problem which is
%\begin{equation}\label{op[13}
%    \int_{-\infty}^{+\infty}e^{-\beta x^2-\gamma x}\cos bx dx=\sqrt{\frac{\pi}{\beta}}e^{\frac{\gamma^2-b^2}{4\beta}}\cos(\frac{b\gamma}{2\beta})\ \ \ \mbox{Re}\{\beta\}>0.
%\end{equation}
%Then (by substituting $\beta=\frac{\lambda^2-1}{2},\ \gamma=j\omega-\lambda y_1,\ b=y_2$) our problem would be equivalent to (noting that $\cosh jx=\cos x$)
%\begin{equation}\label{op14}
%    e^{-\frac{\omega^2}{2(\lambda^2-1)}}\int_{-\infty}^{\infty}\!\!\int_{-\infty}^{\infty}e^{\frac{y_1^2}{2(\lambda^2-1)}-\frac{\lambda^2y_2^2}{2(\lambda^2-1)}-j\frac{\lambda y_1}{\lambda^2-1}\omega}\cos \left(\frac{y_2(j\omega-\lambda y_1)}{(\lambda^2-1)}\right)\ln f_{\mathbf Y}(\mathbf y;f^*_{X_1})d\mathbf y=\sqrt{\frac{\lambda^2-1}{2\pi}}c\delta(\omega)
%\end{equation}
%\color{black}

\section{Analysis in polar coordinates}\label{polar}
In this section, the problem in (\ref{4e2}) is analyzed in polar coordinates. Also, in the numerical section, we adopt the notations used in this section. By writing the input and output of the channel in polar coordinates, we have
\begin{equation}\label{4e3}
    \mathbf{X}=R[\cos\Theta\ ,\ \sin\Theta]^T\ \ ,\ \ \mathbf{Y}=P[\cos\Psi\ ,\ \sin\Psi]^T\ \ \ \ \Theta,\Psi\in[0,2\pi)\ ,P\in[0,\infty).
\end{equation}
Therefore,
\begin{align}
    h(\mathbf Y)&=-\int_{\mathbb{R}^2}f_{\mathbf Y}(\mathbf y)\ln f_{\mathbf Y}(\mathbf y)d\mathbf y \label{e5}\\
    &=-\int_{0}^{\infty}\!\!\int_{0}^{2\pi}f_{\mathbf Y}(\mathbf y(\rho, \psi))\ln f_{\mathbf Y}(\mathbf y(\rho, \psi))|\frac{\partial \mathbf y}{\partial(\rho, \psi)}|d\psi d\rho\\
    &=-\int_{0}^{\infty}\!\!\int_{0}^{2\pi}f_{P, \Psi}(\rho, \psi)\ln \frac{f_{P, \Psi}(\rho, \psi)}{|\frac{\partial \mathbf y}{\partial(\rho, \psi)}|}d\psi d\rho\\
    &=h(P,{\Psi})+\int_{0}^{\infty}f_P(\rho)\ln \rho d\rho\label{e6}\\
    &=h(V,\Psi)
\end{align}
where $\frac{\partial \mathbf y}{\partial(\rho, \psi)}=\rho$ is the Jacobian of the transform and $V=\frac{P^2}{2}$. It can be easily verified that
\begin{equation}\label{4e6}
    f_{V,\Psi}(v,\psi;F_\Theta)=\int_{0}^{2\pi}\tilde{K}(v,\psi,\theta)dF_{\Theta}(\theta)
\end{equation}
where the kernel function is given by
\begin{equation}\label{4e7}
    \tilde{K}(v,\psi,\theta)=\frac{1}{2\pi}e^{-\frac{\lambda^2-1}{2}R^2\cos^2\theta-\frac{R^2}{2}+R\sqrt{2v}(\lambda\cos\psi\cos\theta+\sin\psi\sin\theta)-v}.
\end{equation}
The marginal entropy density of the output variables induced by the input distribution is defined as
\begin{equation}\label{4e8}
    \tilde{h}_{V,\Psi}(\theta;F_{\Theta})=-\int_{0}^{\infty}\!\!\int_{0}^{2\pi}\tilde{K}(v,\psi,\theta)\ln f_{V,\Psi}(v,\psi;F_\Theta)d\psi dv
\end{equation}
which satisfies the following
\begin{equation}\label{4e9}
    h(V,\Psi;F_\Theta)=\int_{0}^{2\pi}\tilde{h}_{V,\Psi}(\theta;F_{\Theta})dF_{\Theta}(\theta).
\end{equation}
Finally, the optimization problem in (\ref{4e2}) becomes
\begin{equation}\label{4e5}
    C(R) = \sup_{F_{\Theta}(\theta)} h(V,\Psi;F_\Theta)-\ln (2\pi e)
\end{equation}

Let $\epsilon^*_\theta$ denote the set of points of increase for the optimal input phase distribution $F^*_\Theta(\theta)$. Analogous to the proof of the theorem in Cartesian coordinates, Lagrangian optimization gives the necessary and sufficient condition for the unique maximizer $F^*_{\Theta}(\theta)$ as
\begin{align}
    \tilde{h}_{V,\Psi}(\theta;F^*_{\Theta})&=h(V,\Psi;F^*_{\Theta})\ \ \ \forall \theta \in \epsilon^*_\theta\nonumber\\
    \tilde{h}_{V,\Psi}(\theta;F^*_{\Theta})&<h(V,\Psi;F^*_{\Theta})\ \ \ \forall \theta\in [0,2\pi)-\epsilon^*_\theta.\label{4e11}
\end{align}
%Further, $|\epsilon^*_\theta|<\infty$.
For the second part of the proof (i.e., showing that $|\epsilon^*_\theta|<\infty$), the difference between investigating the problem in the Cartesian and polar coordinates is in the extension to complex domain. In other words, the kernel and marginal entropy density are entire functions (i.e., holomorphic on the whole complex plane) in polar coordinates. This helps us avoid the consideration of checking the position of accumulation point (see the paragraph below (\ref{op7})). Therefore, the assumption of an infinite number of points of increase results in
\begin{equation}\label{4e12}
    \tilde{h}_{V,\Psi}(z;F^*_{\Theta})=h(V,\Psi;F^*_{\Theta})\ \ ,\ \ \forall z \in \mathbb{C}
\end{equation}
or equivalently
\begin{equation}\label{4e13}
    -\frac{1}{2\pi}e^{-\frac{R^2}{2}}\int_{0}^{\infty}\!\!\!\int_{0}^{2\pi}\!\!e^{-\frac{\lambda^2-1}{2}R^2\cos^2z+R\sqrt{2v}(\lambda\cos\psi\cos z+\sin\psi\sin z)-v}\ln f_{V,\Psi}(v,\psi;F^*_{\Theta})d\psi dv=c\ \ ,\ \ \forall z \in \mathbb{C}
\end{equation}
where $c$ is a constant ($=h(V,\Psi;F^*_{\Theta})$). %For $z(=x+iy)$, the following identities hold
%\begin{align}
%    \cos z &= \cos x\cosh  y -i\sin x\sinh y \\
%    \sin z &= \sin x\cosh y +i\cos x\sinh y.
%\end{align}
Taking $z$ on the imaginary line, we have
\begin{align}\label{sin}
    \cos z = t\ (t\geq1)\ \ ,\ \     \sin z =i\sqrt{t^2-1}.
\end{align}
By replacing (\ref{sin}) in (\ref{4e13}), we get
\begin{equation}\label{4e15}
    -\frac{1}{2\pi}e^{-\frac{R^2}{2}}\int_{0}^{\infty}\!\!\!\int_{0}^{2\pi}\!\!e^{-\frac{\lambda^2-1}{2}R^2t^2+R\sqrt{2v}(\lambda\cos\psi t+i\sin\psi\sqrt{t^2-1})-v}\ln f_{V,\Psi}(v,\psi;F^*_{\Theta})d\psi dv=c\ ,\ \forall t\geq 1.
\end{equation}
Finally, by separating the real and imaginary parts of the left-hand side of (\ref{4e15}), the following is resulted
\begin{align}
     -\frac{1}{2\pi}e^{-\frac{R^2}{2}}\int_{0}^{\infty}\!\!\!\int_{0}^{2\pi}\!\!e^{-\frac{\lambda^2-1}{2}R^2t^2+R\sqrt{2v}\lambda\cos\psi t-v}\cos \left(\sin\psi R\sqrt{2v(t^2-1)}\right)\ln f_{V,\Psi}(v,\psi;F^*_{\Theta})d\psi dv&=c  \label{4e16}\\
     -\frac{1}{2\pi}e^{-\frac{R^2}{2}}\int_{0}^{\infty}\!\!\!\int_{0}^{2\pi}\!\!e^{-\frac{\lambda^2-1}{2}R^2t^2+R\sqrt{2v}\lambda\cos\psi t-v}\sin \left(\sin\psi R\sqrt{2v(t^2-1)}\right)\ln f_{V,\Psi}(v,\psi;F^*_{\Theta})d\psi dv&=0.  \label{4e17}
\end{align}
It is easy to verify that the integrand of (\ref{4e17}) is an odd function with respect to $\psi=\pi$ which is a consequence of the symmetry of the additive noise. Therefore, (\ref{4e17}) is always true. The way to show that (\ref{4e16}) does not hold is similar to that for disproving (\ref{op10}).

%--------------------------------------------

\section{Numerical results}\label{nr}
The theorem in section \ref{th} states that the optimal input has a finite number of mass points on the circle defined by the constraint. The algorithm for finding the number, the positions and the probabilities of these points is the same as that explained in \cite{Shamai} where we start with two points for very small $R$ and then increase $R$ by some step and check the necessary and sufficient conditions. At any stage that these conditions are violated, we increase the number of points, do the optimization to find the position and probabilities of the points, check the conditions and keep repeating this process.

\begin{figure}[H]
\centering
\begin{subfigure}{.5\textwidth}
  \centering
  \includegraphics[width=8 cm]{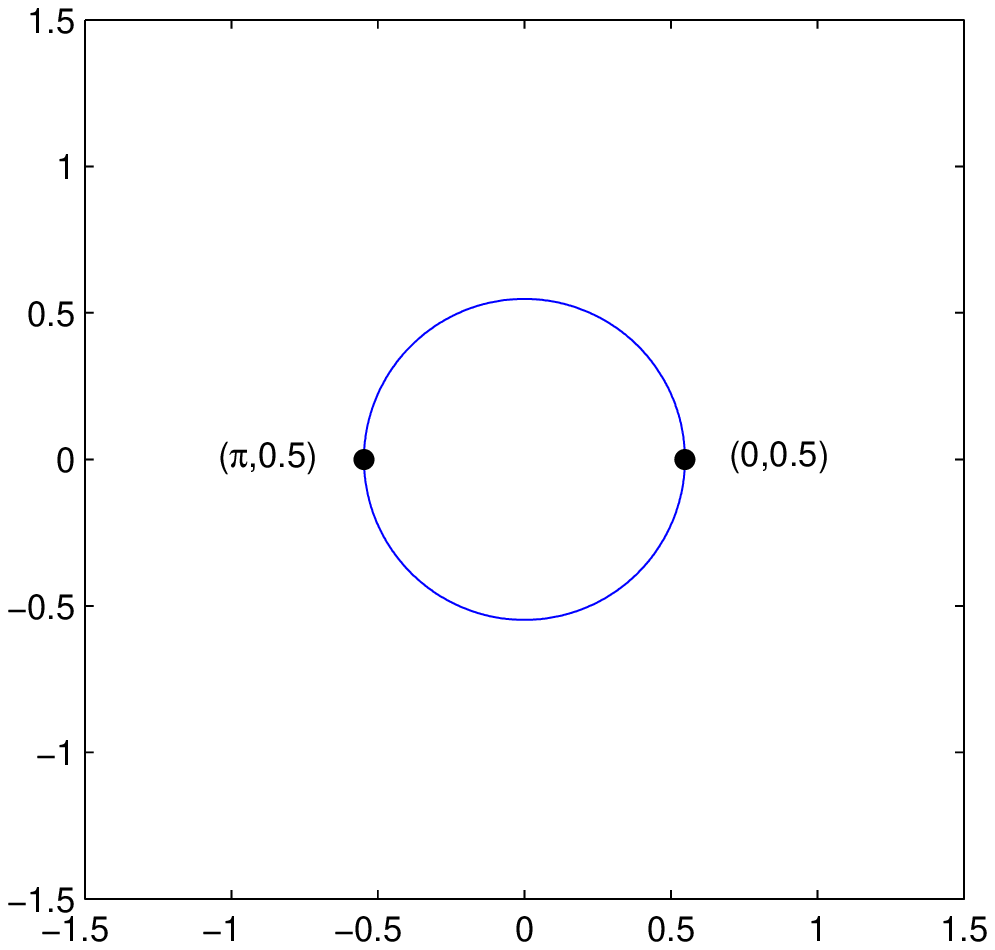}
  \caption{}
  \label{fig1:sub1}
\end{subfigure}%
\begin{subfigure}{.5\textwidth}
  \centering
  \includegraphics[width=8 cm]{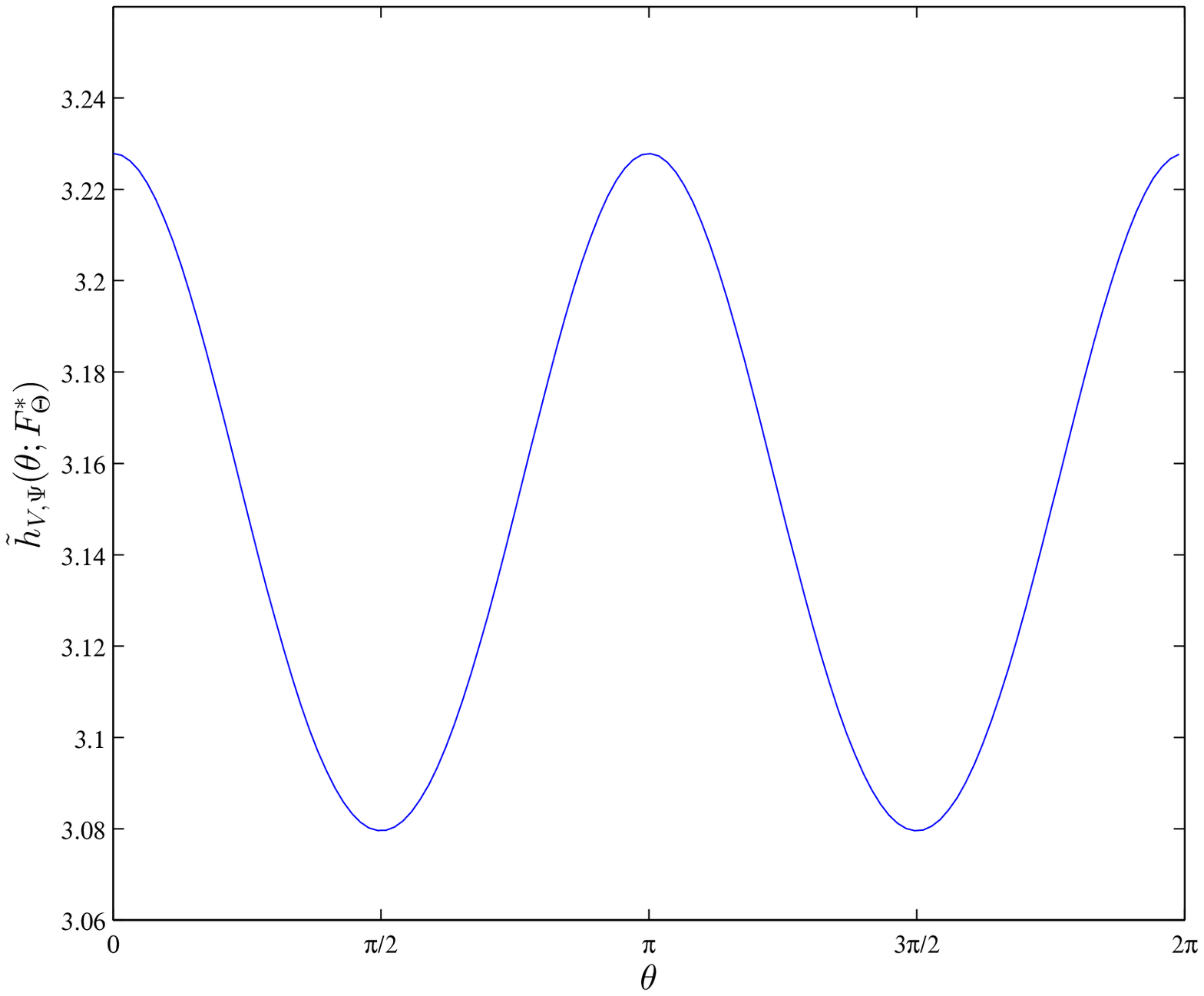}
  \caption{}
  \label{fig1:sub2}
\end{subfigure}
\caption{The support of the optimal input (a) and the marginal entropy density induced by it (b) for $R=0.5477$ and $\lambda = 2$. In (a), the pairs represent the phase and its probability as in $(\theta,P_{\Theta}(\theta))$.}
\label{ffig1}
\end{figure}

\begin{figure}[H]
\centering
\begin{subfigure}{.5\textwidth}
  \centering
  \includegraphics[width=8 cm]{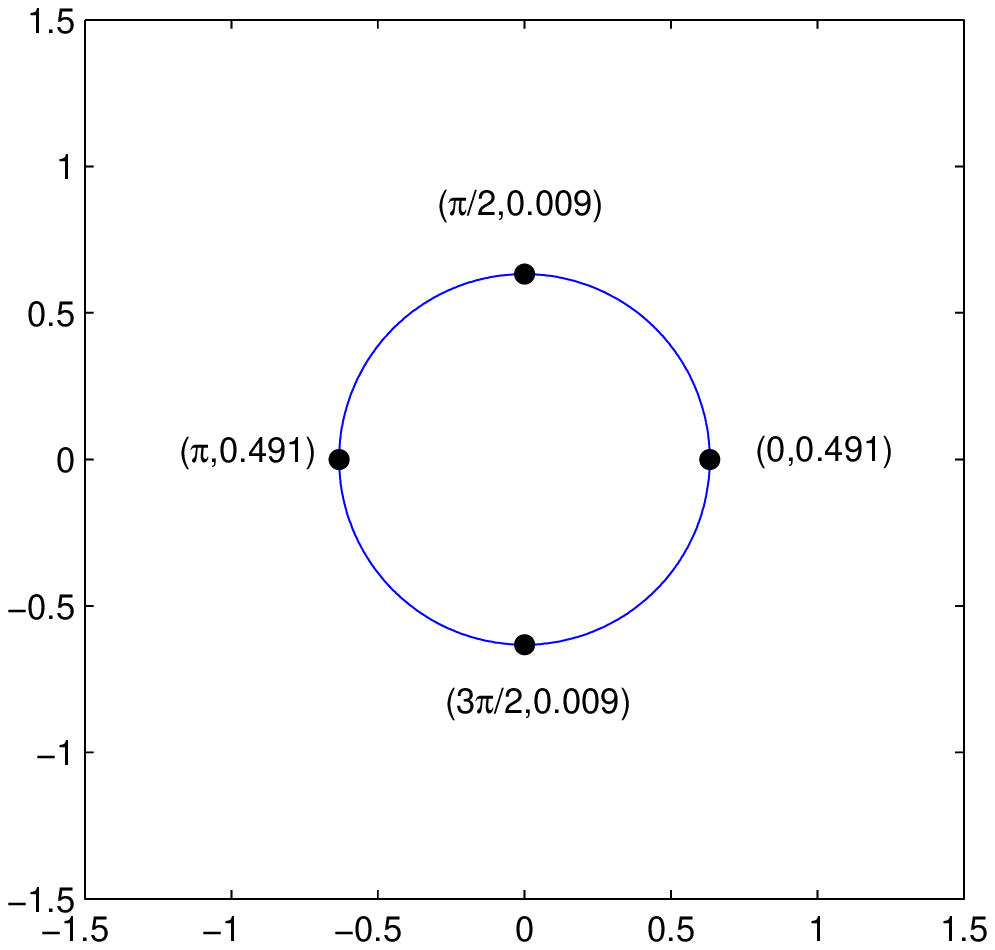}
  \caption{}
  \label{fig2:sub1}
\end{subfigure}%
\begin{subfigure}{.5\textwidth}
  \centering
  \includegraphics[width=8 cm]{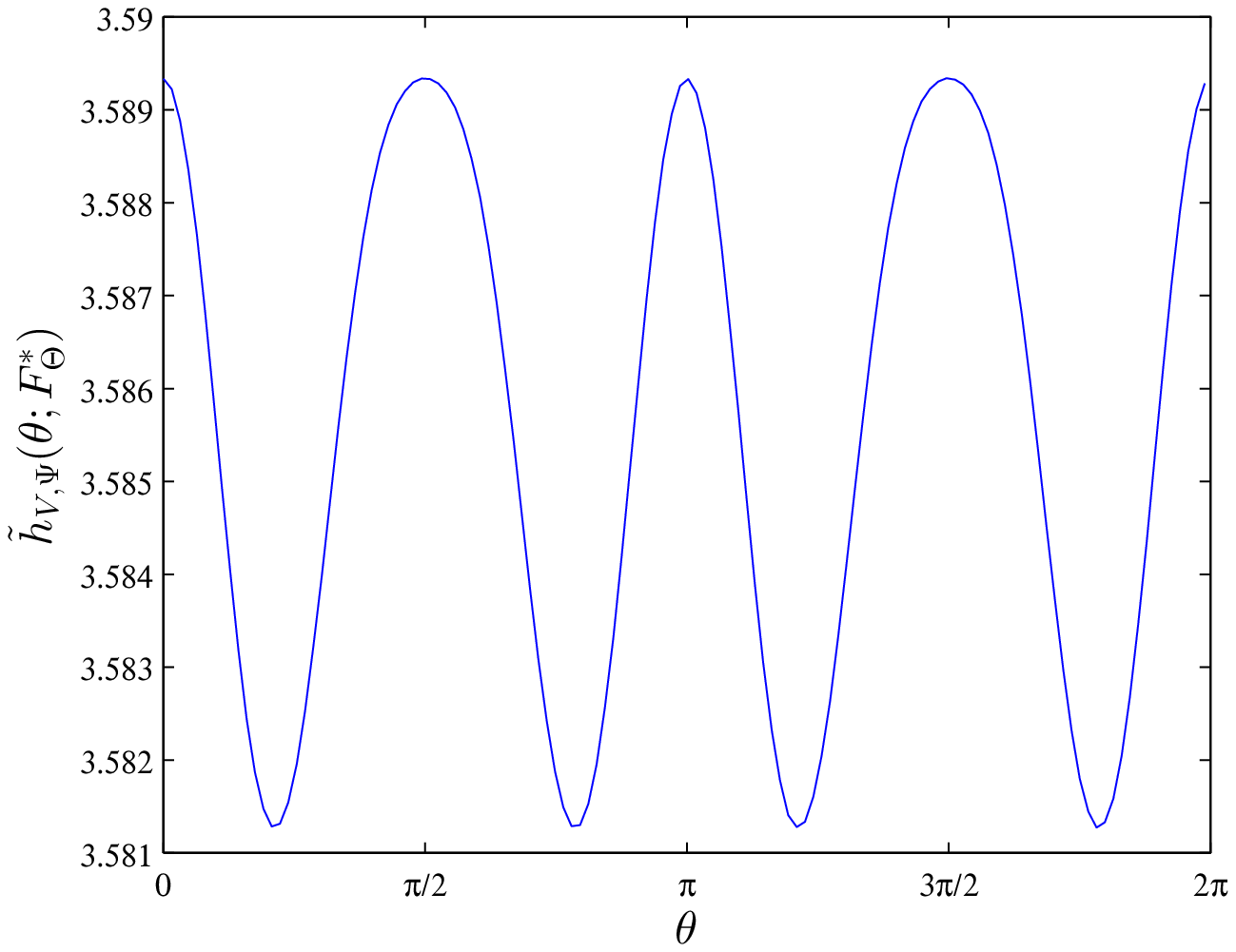}
  \caption{}
  \label{fig2:sub2}
\end{subfigure}
\caption{The support of the optimal input (a) and the marginal entropy density induced by it (b) for $R=0.6325$ and $\lambda = 2$. In (a), the pairs represent the phase and its probability as in $(\theta,P_{\Theta}(\theta))$.}
\label{ffig2}
\end{figure}

\begin{figure}[H]
\centering
\begin{subfigure}{0.5\textwidth}
  \centering
  \includegraphics[width=8 cm]{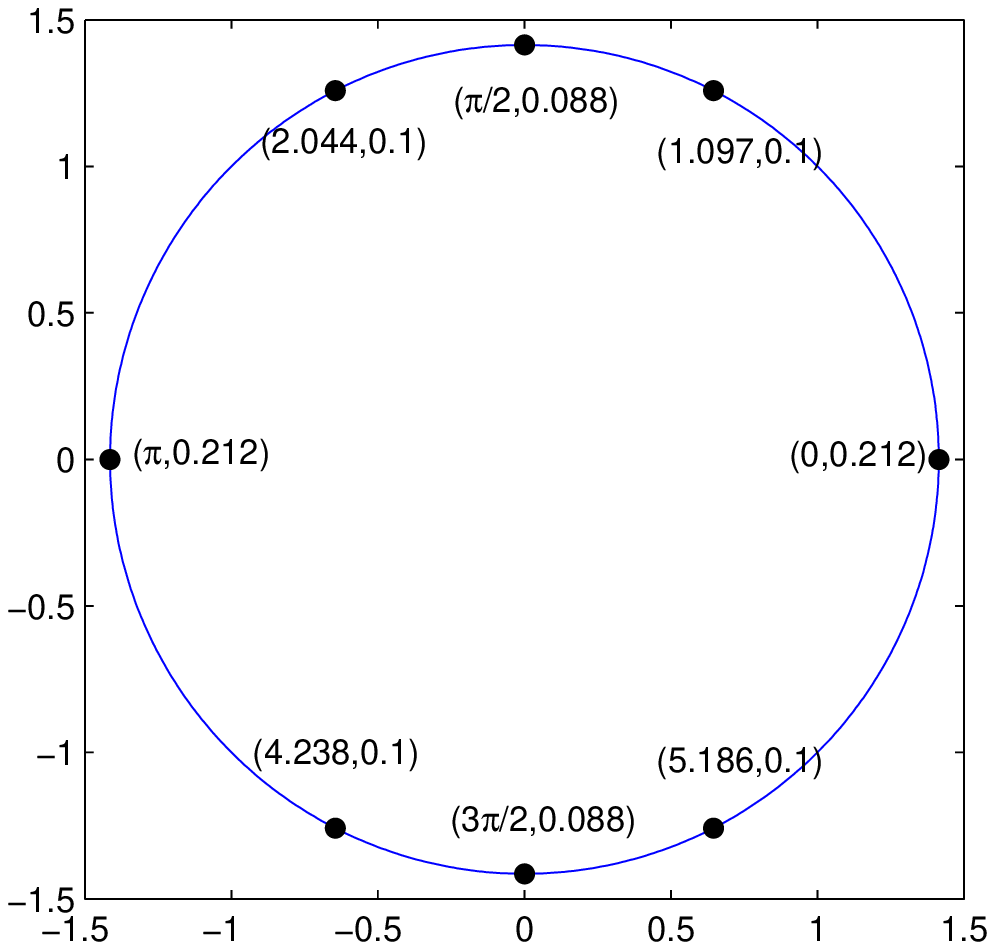}
  \caption{}
  \label{fig4:sub1}
\end{subfigure}%
\begin{subfigure}{0.5\textwidth}
  \centering
  \includegraphics[width=8 cm]{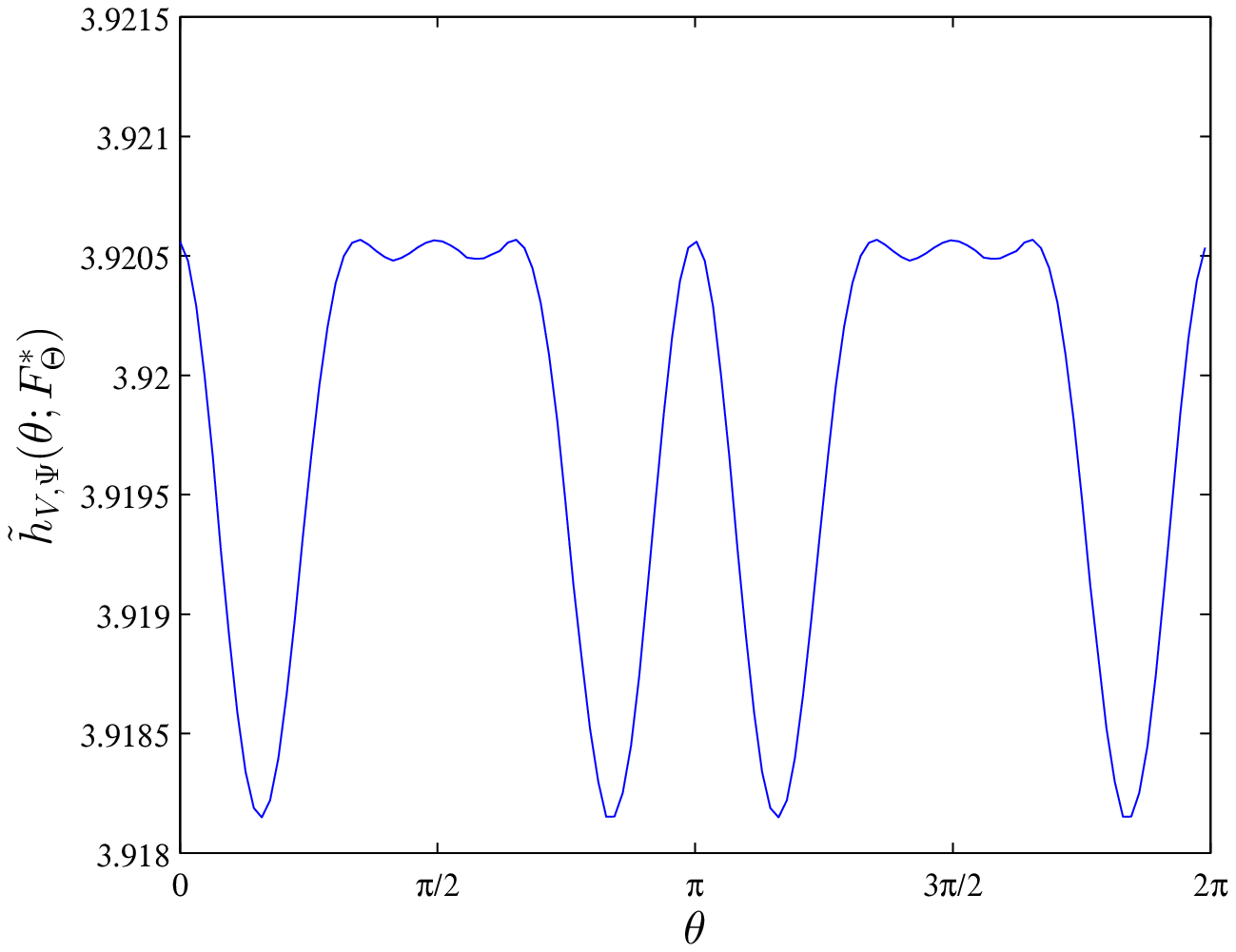}
  \caption{}
  \label{fig4:sub2}
\end{subfigure}
\caption{The support of the optimal input (a) and the marginal entropy density induced by it (b) for $R=\sqrt{2}$ and $\lambda = 2$. In (a), the pairs represent the phase and its probability as in $(\theta,P_{\Theta}(\theta))$.}
\label{ffig4}
\end{figure}

The support of the capacity achieving input and the marginal entropy densities induced by them are shown in figures \ref{ffig1} to \ref{ffig4} for $\lambda=2$ and different values of $R$. Here, we have performed the optimization in polar coordinates. The optimality of the points in the left figures (i.e., figures \ref{fig1:sub1} to \ref{fig4:sub1}) is guaranteed by the necessary and sufficient conditions in (\ref{4e11}) which can also be verified through figures \ref{fig1:sub2} to \ref{fig4:sub2}. As it can be observed, the points of increase of the optimal input, which correspond to the peaks in the marginal entropy densities, have a finite number.

Let $F^1_{X_1}(x)$ be defined as
\begin{equation}\label{n1}
    F^1_{X_1}(x)=\frac{1}{2}\left[u(x-R)+u(x+R)\right].
\end{equation}
According to section \ref{asymbeh}, we know that this CDF is optimal for sufficiently small values of $R$. As $R$ increases, $F^1_{X_1}(x)$ remains optimal until it violates the necessary and sufficient conditions.
By observing the behavior of $\tilde{h}_{\mathbf Y}(x;F^1_{X_1})$, it is concluded
that as $R$ increases, the first point to violate the necessary and sufficient conditions will happen at $x=0$ (which is equivalent to $(0,R)$ and $(0,-R)$ on the circle).

%\begin{figure}[H]
%\centering
%\begin{subfigure}{.5\textwidth}
%  \centering
%  \includegraphics[width=8 cm]{fig03}
%  \caption{}
%  \label{fig3:sub1}
%\end{subfigure}%
%\begin{subfigure}{.5\textwidth}
%  \centering
%  \includegraphics[width=8 cm]{FFF}
%  \caption{}
%  \label{fig3:sub2}
%\end{subfigure}
%\caption{The support of the optimal input (a) and the marginal entropy density induced by it (b) for $R=1.0954$ and $\lambda = 2$. In (a), the pairs represent the phase and its probability as in $(\theta,P_{\Theta}(\theta))$.}
%\label{ffig3}
%\end{figure}

\begin{figure}[H]
  \centering
  \includegraphics[width=10cm]{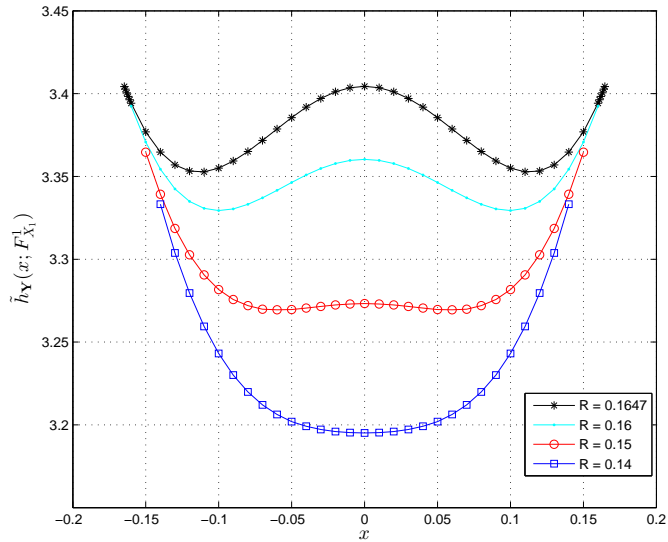}\\
  \caption{For small values of $R$, when $R$ increases, the first point to become a mass point is $x=0$. (here $\lambda = 10$).}\label{zero point}
\end{figure}

 This is shown
in figure \ref{zero point} for $\lambda = 10$. Therefore, the norm threshold ($R^t$) for which $F^1_{X_1}(x)$ remains optimal is obtained by solving the following equation for $R^t$
\begin{equation}\label{n2}
    \tilde{h}_{\mathbf Y}(0;F^1_{X_1})=\tilde{h}_{\mathbf Y}(R;F^1_{X_1})
\end{equation}
which, after some manipulation, becomes equivalent to
\begin{equation}\label{n3}
    \frac{1}{\sqrt{2\pi}}\int_{-\infty}^{+\infty}(e^{-\frac{(y-\lambda R)^2}{2}}-e^{-\frac{y^2}{2}})\ln\cosh(\lambda Ry)dy=\frac{(\lambda^2-1)}{2}R^2.
\end{equation}

\begin{figure}[H]
  \centering
  \includegraphics[width=10cm]{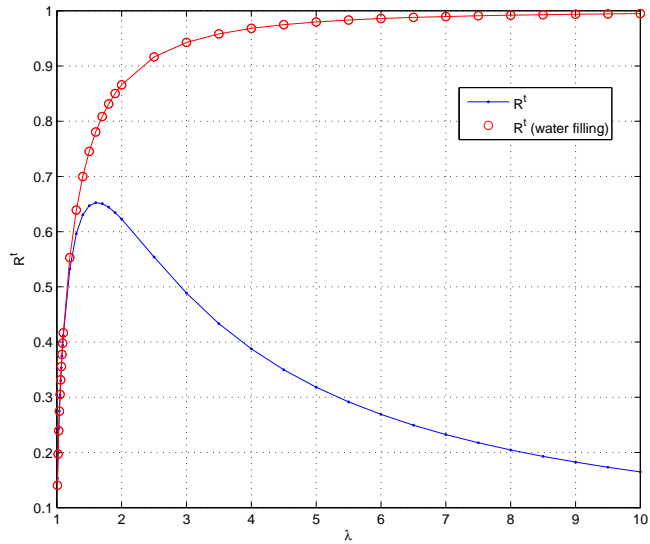}\\
  \caption{The value of the norm threshold as a function of $\lambda$.}\label{threshold}
\end{figure}

\begin{figure}[H]
\centering
\begin{subfigure}{0.33\textwidth}
  \centering
  \includegraphics[width=6 cm]{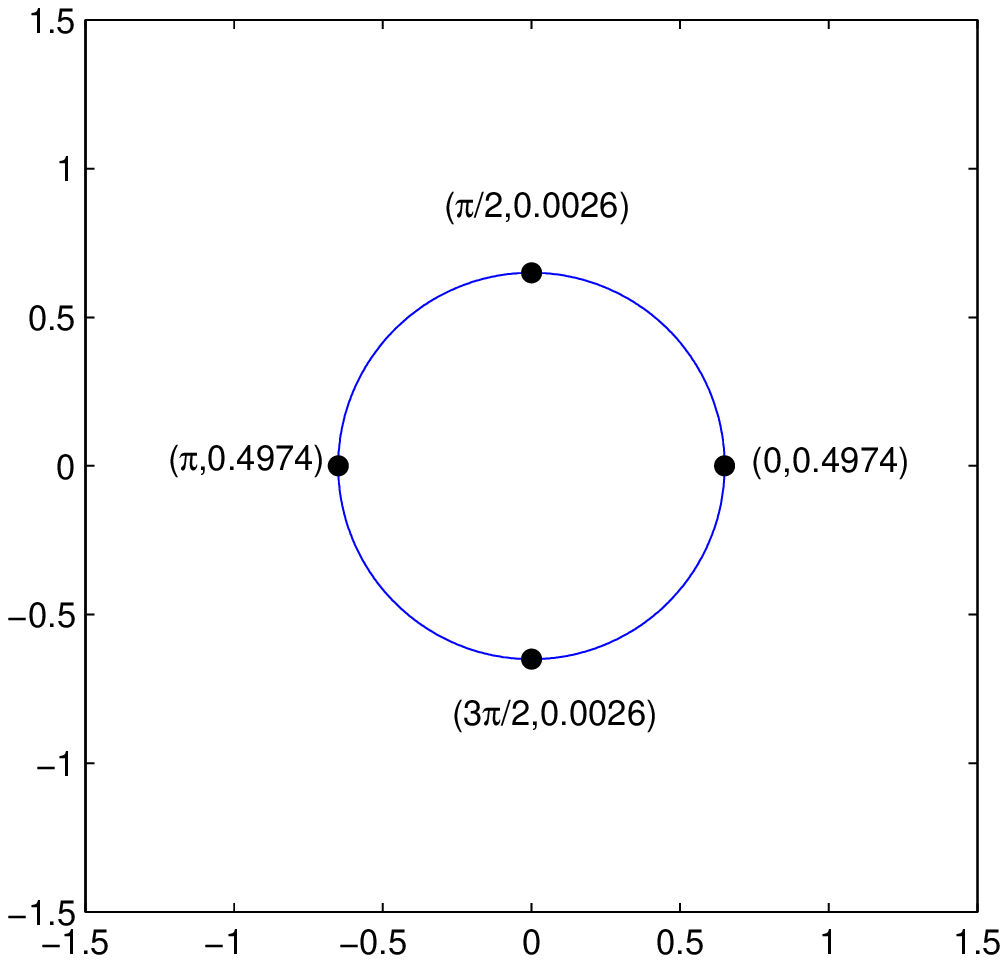}
  \caption{$\lambda=1.5$}
  \label{fig5:sub1}
\end{subfigure}%
\begin{subfigure}{0.33\textwidth}
  \centering
  \includegraphics[width=6 cm]{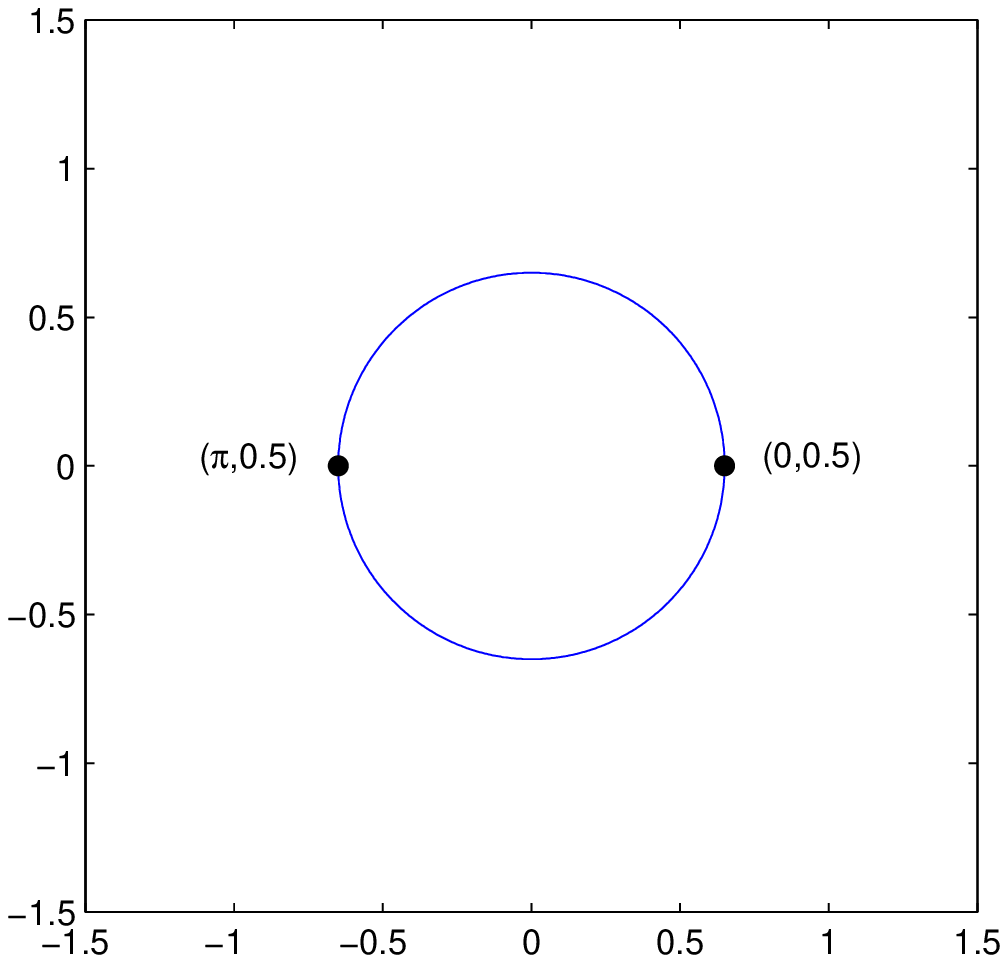}
  \caption{$\lambda=1.6$}
  \label{fig5:sub2}
\end{subfigure}%
\begin{subfigure}{0.33\textwidth}
  \centering
  \includegraphics[width=6 cm]{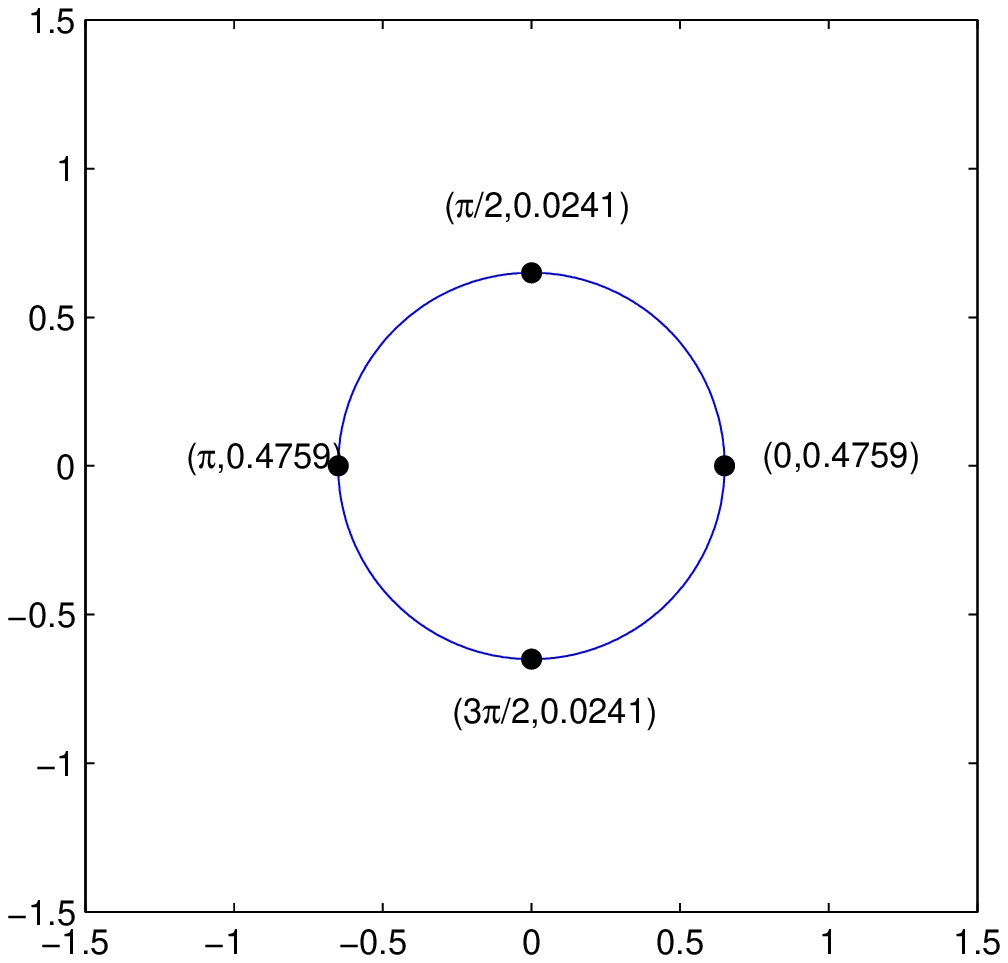}
  \caption{$\lambda=2$}
  \label{fig5:sub3}
\end{subfigure}
\caption{The support of the optimal input for $R=0.65$ and different values of $\lambda$.}
\label{ffig5}
\end{figure}

By solving (\ref{n3}) numerically, the values of $R^t$ are obtained for different values of $\lambda$ as shown in figure \ref{threshold}.
For example, for $\lambda=10$, $R^t=0.1647$ which means that when the norm $R$ is below $0.1647$ the
support of the optimal input has only two equiprobable mass points at $(R,0)$ and $(-R,0)$, and at this threshold it gets another mass point at zero as already shown in figure \ref{zero point}. In the conventional case (i.e., the case with only average power constraint), according to (\ref{as2.2}) and (\ref{as2}), the water filling algorithm starts allocating power to $X_2$ at $R=\sqrt{1-\frac{1}{\lambda^2}}$. The red curve in figure \ref{threshold} shows this water level which is the conventional counterpart of $R^t$ for the constant envelope signaling. As it can be observed both curves have the same trend when the channel is ill-conditioned, but diverge as the condition number increases. In the conventional case, when $R<1$, by increasing the quality of the stronger channel (i.e., by increasing $\lambda$), the weaker channel will become inactive, while in the constant envelope signaling, increasing the quality of the stronger channel does not have a similar result. To elaborate this further, figure \ref{ffig5} shows an example in which $R=0.65$ and by increasing $\lambda$, the weaker channel becomes inactive first and then active. This is due to the behavior of $R^t$ in figure \ref{threshold} which first increases and then decreases. It can be verified that for $R=0.65$, the distribution in (\ref{n1}) is optimal only in the interval $\lambda\in(1.55,1.7)$.

\section{Conclusion}\label{conc}
In this paper, constant envelope signaling in point-to-point Gaussian MIMO channels was considered. For a 2 by 2 channel, we showed that the capacity-achieving input distribution has a finite number of mass points on the circle defined by the constant norm. In this setting, the optimal DoF of a full rank $n$ by $n$ channel was shown to be $n-1$ which is achieved by a uniform distribution over the surface of the hypersphere defined by the constant envelope. Finally, when the channel is ill-conditioned, the performance of the constant envelope signaling was shown to be similar to that of the conventional case (i.e., water-filling) for a 2 by 2 channel in terms of power allocation. However, it was observed that this similarity vanishes as the condition number grows.
\bibliography{REFERENCE}
\bibliographystyle{IEEEtran}
\end{document}